\documentstyle[prc,aps,twocolumn]{revtex}

\begin{document}

\draft

\input{psfig}

\twocolumn[
\hsize\textwidth\columnwidth\hsize\csname @twocolumnfalse\endcsname

\title{
\hfill {\small Preprint TRI-PP-00-29} \\
Photonuclear Reactions of Three-Nucleon Systems}

\author{W.~Schadow$^1$, O. Nohadani$^2$, W.~Sandhas$^2$}

\address{
$^1$TRIUMF, 4004 Wesbrook Mall, Vancouver, British Columbia, Canada V6T 2A3}
\address{$^2$Physikalisches  Institut der Universit\"at Bonn, Endenicher
Allee 11-13, D-53115 Bonn, Germany}

\vspace{6mm}

\date{\today}

\maketitle

\begin{abstract}
We discuss the available data for the differential and the total cross
section for the photodisintegration of $^3$He and $^3$H and the
corresponding inverse reactions below $E_\gamma = 100$ MeV by
comparing with our calculations using realistic $NN$ interactions. The
theoretical results agree within the errorbars with the data for the
total cross sections. Excellent agreement is achieved for the angular
distribution in case of $^3$He, whereas for $^3$H a discrepancy
between theory and experiment is found.
\end{abstract}

\vspace{-2pt}

\pacs{PACS numbers(s): 21.45.+v, 25.40.Lw, 25.20.-x, 27.10.+h}

\vspace{6pt}

]

\section{Introduction}
Over the last decades the photodisintegration of $^3$He and $^3$H and
the corresponding inverse reactions have been investigated
experimentally and theoretically with considerable interest. There
have been a lot of experiments using different techniques for the
photodisintegration of $^3$He
\cite{Griffith62a,Stewart63,Warren63a,Fetisov65a,Berman64a,Woelfli66,Woelfli67,Belt70,Kundu71a,Ticcioni73,Chang74,Skopik79a,King84a,King84b,Vanderwoude71a,OFallon72a,Chang72a,Matthews74,Anghinolfi83a,Skopik83a,Pitts88a}
and $^3$H
\cite{Boesch64,Kosiek66,Pfeiffer68,Faul80,Skopik81,Mitev86a,Moesner86a}
or the inverse reaction, respectively. Despite the many
investigations, there are inconsistencies between the data up to 30\%
in the magnitudes of the cross sections.

Early theoretical calculations were restricted to phenomenological
interactions and various approximations in the bound state wave
function and the scattering states (for a list of References see
\cite{Klepacki88b}). The first consistent calculation for both the
initial and the final state was done by Gibson and Lehman
\cite{Gibson75}. They solved Faddeev-type Alt-Grassberger-Sandhas
(AGS) equations \cite{Alt67} using Yamaguchi interactions and taking
into account only the $E1$ contributions of the electromagnetic
interaction.

Attempts to use realistic interactions are the ones by Aufleger and
Drechsel \cite{Aufleger81a} and by Craver {\em et
al}. \cite{Craver77a}. In both calculations higher multipoles were
considered, but the three-body scattering state was not treated
exactly. The unusual energy dependence of the cross section postulated
by Craver {\em et al}. has never been confirmed by any other
calculation.  Klepacki {\em et al}. \cite{Klepacki88b} also used a
realistic interaction, however, in plane wave impulse
approximation. King {\em et al}. \cite{King84a} performed an effective
two-body direct capture calculation with the initial state being
treated as a plane wave, or as a scattering state generated from an
optical potential.

The very-low-energy $n$-$d$ radiative capture process is dominated by
the magnetic dipole ($M1$) transition, and has been studied by several
authors \cite{Friar90,Viviani96a} in configuration-space with
inclusion of three-body forces, final state interaction (FSI), and
explicit meson exchange currents (MEC).  The inclusive reaction (two-
and three-fragment) has been studied recently by Efros {\em et
al}. \cite{Efros99a} with realistic two-body interactions and
three-body forces as input in the energy range up to 100 MeV by
employing the Lorentz integral transformation method.  Other recent
theoretical work was devoted to polarization observables for $p$-$d$
capture \cite{Jourdan86,Ishikawa92b,Fonseca95a,Anklin98a,Golak2000}
using realistic interactions. A discussion of polarization observables
will be published in a subsequent paper.

In Refs. \cite{Schadow98a,Sandhas98a,Schadow99a}, we have treated the
$^3$He and $^3$H photodisintegration and the inverse processes, i.e.,
the radiative capture of protons or neutrons by deuterons, within the
integral equation approach discussed below.  These calculations were
based on the Bonn {\em A}, Bonn {\em B} and Paris potentials in
Ernst-Shakin-Thaler (EST) expansion: Bonn {\em A } (EST), Bonn {\em B}
(EST), and Paris (EST) \cite{Haidenbauer84,Haidenbauer86a}. We have
demonstrated, in particular, the role of $E2$ contributions, meson
exchange currents, and higher partial waves. A noticeable potential
dependence was found in the peak region, i.e., for $E_x \leq 20$ MeV.
But it was also shown that the different peak heights are strongly
correlated with the different three-nucleon binding energies obtained
for the potentials employed. The possibility of using the magnitudes
of the cross sections as an independent test of the quality of the
potentials, thus, appears rather restricted.

The aim of the present paper is to extend the investigations of
Refs. \cite{Schadow98a,Sandhas98a,Schadow99a} to photon energies from
20~MeV to 100~MeV. Aside from certain energy points there are up to
now no other theoretical calculations for the differential and the
total cross section available in this energy range using the the full
final-state amplitudes, realistic interactions and taking into account
$E1$ and $E2$ contributions of the electromagnetic interaction.

This paper is organized as follows: In section \ref{sectheory} we
present the theoretical framework of our calculations. The results are
discussed is section \ref{secresults}. Our conclusions are summarized
in section \ref{secconclusions}.

\section{Theory}
\label{sectheory}

The amplitude for the two-fragment photodisintegration of $^3$H or
$^3$He into a deuteron $\psi_{d}$ and a neutron or proton of relative
momentum ${\bf q}$ is given by

\begin{eqnarray}
\label{bampl}
 M ({\bf q}\,) &=& \; ^{(-)}_{\;\;\;\cal S}\langle{\bf q}; \psi_{d}|
H_{\text{\scriptsize em}}|\Psi_{\text {BS}}
\rangle_{\cal S} \nonumber \\
&=& \; ^{(-)}_{\;\;\;\cal S}\langle \Psi|
H_{\text{\scriptsize em}}|\Psi_{\text {BS}}
\rangle^{}_{\cal S} .
\end{eqnarray}

\noindent
Here $|\Psi_{\text {BS}} \rangle^{}_{\cal S}$ represents the incoming
three-nucleon bound state while $^{(-)}_{\;\;\;\cal S}\langle{\bf q};
\psi_{d}|$ denotes the final continuum (scattering) state with
outgoing boundary condition.  $H_{\rm em}$ is the electromagnetic
operator.  The initial and final states are assumed to be properly
antisymmetrized. The antisymmetrized final state can be represented as
a sum over the three possible two-fragment partitions $\beta$

\begin{equation}
^{(-)}_{\;\;\;\cal S}\langle{\bf q}; \psi_{d}| = \;
^{(-)}_{\;\;\;\cal S}\langle \Psi| =
\frac{1}{\sqrt{3}} \sum_\beta\; ^{(-)}\langle \Psi_\beta| \, .
\end{equation}

\noindent
The scattering state $^{(-)}\langle{\Psi_\beta}|$ can be obtained from
the free channel state $\langle \Phi_\beta |$ via the M{\o}ller
operators $\Omega_\beta^{(-)}$, i.e., $^{(-)}\langle{\Psi_\beta}| =
\langle {\Phi_\beta}| \, \Omega_\beta^{(-) \dagger}$. The channel
state

\begin{equation}
\langle{\Phi_\beta}| = \langle{(\beta) {\bf q} \, S M_S; \tau M_\tau}| \,
\langle{(\beta) \Psi^\eta  m_j; m_t}|
\label{eqchannelstate}
\end{equation}

\noindent
is a tensor product of the deuteron wave function and a plane wave
state of the outgoing third particle.  Moreover, it is an eigenstate
of the channel Hamiltonian $H_\beta = H_0 + V_\beta$.  Here we have
used the complementary notation $V_\beta = V_{\gamma \alpha}$ for the
two-body potentials, i.e., $V_{\gamma \alpha}$ denotes the interaction
between the particle $\gamma$ and $\alpha$, while $H_0$ denotes the
free three-body Hamiltonian. The states in Eq. (\ref{eqchannelstate}) are
labeled by the corresponding quantum numbers. The collective index
$\eta = (s \, j;t)$ stands for the spin $s$, the total angular momentum
$j$, and the isospin $t$ of the two-body subsystem. The indices $S$
and $\tau$ denote the spin and isospin of the third particle,
respectively.\par

It can be shown \cite{Sandhas86a} that the adjoint M{\o}ller operators
satisfy the relation

\begin{equation}
\label{moeller1}
  \Omega_\beta^{(-) \dagger} = \delta_{  \beta \alpha} + U_{ \beta \alpha}
(E_\beta + i0) \, G_\alpha (E_\beta + i0) \,,
\end{equation}

\noindent
where $U_{ \beta \alpha}$ are the usual AGS \cite{Alt67} operators,
and $G_\alpha$ is the resolvent of the channel Hamiltonian $H_\alpha =
H_0 + V_\alpha$.

\vfill

\begin{minipage}[t]{8.3cm}
\begin{table}[h]
\begin{tabular}{lcccc}
 partial wave       & Bonn {\em A} (EST) & Bonn {\em B} (EST) & Paris (EST) \\
\hline \vspace{-3mm} \\
 $^1S_0$            &    5   &   5  &   5  \\
 $^3S_1-\,^3\!D_1$  &    6   &   6  &   6  \\
 $^1D_2$            &    4   &   4  &   5  \\
 $^3D_2$            &    4   &   4  &   5  \\
 $^1P_1$            &    4   &   4  &   5  \\
 $^3P_1$            &    4   &   4  &   5  \\
 $^3P_0$            &    4   &   4  &   5  \\
 $^3P_2-\,^3\!F_2$  &    5   &   5  &   7  \\
\end{tabular}

\vspace{3mm}
\caption{Ranks of the two-body partial waves of the Paris, Bonn {\it
A}, and Bonn {\it B} potentials in EST representation used for the
bound-state and the scattering calculations.
\label{ranktab}}
\end{table}
\end{minipage}

\noindent
Multiplying the AGS equations

\begin{equation}
  \label{AGS}
  U_{\beta \alpha} = (1-\delta_{\beta \alpha}) G_0^{-1}
  + \sum_{\gamma} (1-\delta_{\beta \alpha}) T_{\gamma} G_0 U_{\gamma \alpha}
\end{equation}

\noindent
from the right with $G_\alpha$ and adding $\delta_{\beta \alpha}$ on
both sides, we see that the left-hand side is already the expression
(\ref{moeller1}), and inserting this relation on the right-hand side
of (\ref{AGS}) we end up, after some trivial manipulations, with the
set of integral equations
\begin{equation}
\label{moellereq}
 \Omega_\beta^{(-) \dagger} = 1 + \sum_\gamma (1 - \delta_{\beta \gamma})
\,T_\gamma \,G_0 \,\Omega_\gamma^{(-) \dagger} \,
\end{equation}

\noindent
for the adjoint M{\o}ller operators. Applying the M{\o}ller operator
(\ref{moellereq}) onto the state $H_{\text{\scriptsize
em}}|\Psi_{\text {BS}}\rangle$, they go over into a set of effective
two-body equations when representing the input two-body $T$ operator
in separable form.  In order to accomplish this, we use the separable
expansion method proposed by Ernst, Shakin, and Thaler \cite{Ernst73}
for representing a given $NN$ interaction. In this scheme the original
potential is expressed as sum over separable terms

\begin{equation}
 V^{\eta}_{l l'} = \sum_{\mu,\nu=1}^N  | g_{\nu}^{\eta} \, l \rangle \,
\Lambda^{\eta}_{\mu \nu} \, \langle g_{\nu}^{\eta} \, l' |  ,
\label{est2}
\end{equation}

\noindent
where $N$ is the rank of the separable representation, $\Lambda_{\mu
\nu}$ are the coupling strengths, and $| g_{\nu}^{\eta} \, l \rangle $
are the form factors.  Here $l$ and $l'$ are the orbital angular
momenta.  The total angular momentum $j$ is obtained from the coupling
sequence $(l s)j$.  Using this representation for the potential, the
two-body $T$ operator reads

\begin{equation}
\label{eqtmatest}
T^{\eta}_{l l'}(E +i0)  = \sum_{\mu \nu }| { g}_{\mu}^{\eta}  \, l\rangle \,
{ \Delta}_{\mu \nu }^{\eta} (E+i0) \,\langle { g}_{\nu}^{\eta} \, l'|\, ,
\end{equation}

\noindent
with

\begin{equation}
 {\bf \Delta}^{\eta} (E + i0) = \left (({{\bf \Lambda}^{\eta}})^{-1} -
{\bf {\cal G}_{0}}(E + i0) \right )^{-1}
\end{equation}

\noindent
and

\begin{equation}
({\bf {\cal G}}_{0}(E + i0))_{\mu \nu} = \sum_l \,
\langle g_{\mu} \, l |G_0(E + i0) | g_{\nu} \, l \rangle  .
\end{equation}

\noindent
For more details of this construction we refer to
Refs. \cite{Haidenbauer84,Haidenbauer86a}.  The ranks for each partial
wave used in this paper for the bound-state and the scattering
calculations are contained in Tab. \ref{ranktab}. The separable
representation reproduces the correct negative--energy bound--state
pole of the two--body $T$ matrix (for the deuteron quantum numbers
$\eta_d$), if the form factor satisfies the homogeneous equation

\begin{eqnarray}
\sum_{l'} V^{\eta_d}_{l l'} G_0(E_d)|g^{\eta_d}_1 \, l'\rangle =
|g^{\eta_d} _1 \, l\rangle \,.
\label{effeqns}
\end{eqnarray}

\noindent
Then the deuteron wave function is given by

\begin{equation}
|\psi_d\rangle =  \sum_l G_0(E_d) \,|g^{\eta_d}_1 \, l\rangle.
\label{eqdeut}
\end{equation}

\noindent
It should be pointed out, that we have renormalized the form factors
and coupling strengths in the $^3S_1$-$^3D_1$ channel from the
original representation in Eq. (\ref{est2}) in order to give the
correct normalized deuteron wave function according to
Eq. (\ref{eqdeut}). This procedure does not change the potential or
the $T$ matrix, since we shift only a normalization constant from the
form factors into the coupling strengths.

Equation (\ref{moellereq}) will be treated numerically in momentum
space, employing a complete set of partial wave states $| p \, q \,l\,
b\, \Gamma M_\Gamma ;I M_I \rangle$.  The label $b$ denotes the set
$(\eta S K L)$ of quantum numbers, where $K$ and $L$ are the channel
spin of the three nucleons [with the coupling sequence $(j S) K$] and
the relative angular momentum between the two-body subsystem and the
third particle, respectively.  $\Gamma$ is the total angular momentum
following from the coupling sequence $(K L) \Gamma$, the total isospin
$I$ follows from the coupling $(t \, \tau) I$. These states satisfy
the completeness relation
\begin{eqnarray}
\label{complete}
1 &=&  \sum_{ \Gamma M_\Gamma} \sum_{l b} \sum_{I M_I}\int\limits_{0}^{\infty}
\int\limits_{0}^{\infty} dp \, p^2 \, dq \, q^2 \, \\
&& \hspace{1cm} \times \;| p \, q \,l\, b\,
\Gamma M_\Gamma ; I M_I\rangle \langle p \, q \,l\, b\, \Gamma M_\Gamma
;I M_I | . \nonumber
\end{eqnarray}
\noindent
The required antisymmetry under permutation of two particles in the
subsystem can be achieved by choosing only those states which satisfy
the condition $(-)^{l + s + t} = -1$ .

Using the above defined states, the partial-wave decomposition of the
channel state defined in Eq.  (\ref{eqchannelstate}) reads
\begin{eqnarray}
 \langle{\Phi_\beta}| & = &
 \sum_{\Gamma {M_\Gamma}} \sum_{b}  \sum_{M_K M_L} \sum_{I M_I}
Y_{L M_L} (\hat q) \,
\langle j m_j S M_S| K M_K \rangle \, \nonumber \\
& & \times  \;
\langle K M_K L M_L |\Gamma M_\Gamma \rangle
\langle t m_t \tau m_\tau | I M_I \rangle \;
\nonumber \\
&& \times \;
 \langle{(\beta) \, g_1  \, q \,  b \, \Gamma M_\Gamma; I M_I}|
\, G_0 (E_d + {\textstyle \frac{3}{4}} q^2 + i0) \,,
\end{eqnarray}
\noindent
with
\begin{eqnarray}
\lefteqn{ \langle{(\beta) \, g_1 \, q \,  b \, \Gamma M_\Gamma; I M_I}|
  } \nonumber \\
&& \hspace{1cm} = \sum_l\int\limits_{0}^{\infty} \! dp \,  p^2 \, \langle{(\beta) \, p
 \, q \, l \,b\,  \Gamma M_\Gamma;  I M_I}| \;\,
g^{\eta_d}_{1 l} (p)  \,,
\label{eqpartwave}
\end{eqnarray}
\noindent
where we have used Eq. (\ref{eqdeut}) for the representation of the
deuteron wave function. As the quantization axis we have chosen the
direction of the incoming photon. The generalization of
Eq. (\ref{eqpartwave}) for arbitrary rank index $\mu$ and arbitrary
channel quantum numbers is given by
\begin{eqnarray}
\lefteqn{ \langle{(\beta) \, g_\mu \, q \,  b \, \Gamma M_\Gamma; I M_I}|}
\nonumber \\
& & \hspace{1cm} = \sum_l\int\limits_{0}^{\infty} \! dp \,  p^2 \,
 \langle{(\beta) \, p
 \, q \, l \,b\,  \Gamma M_\Gamma;  I M_I}| \;\,
g^{\eta}_{\mu l} (p)  \,.
\label{eqpartwavegeneral}
\end{eqnarray}
\noindent
These states are needed to derive and solve an integral equation for
the final-state amplitudes. From the solution of this equation only
those channels with a deuteron as subsystem, i.e., the channels
from Eq. (\ref{eqpartwave}), are needed for the calculation of
observables for the two-body breakup or the capture process.

Equipped with the above equations we can now derive an integral
equation for the scattering amplitudes.  Multiplying
Eq. (\ref{moellereq}) from the left with $G_0$ and the states of
Eq. (\ref{eqpartwavegeneral}), and from the right with $H_{\text
{em}}| \Psi_{\text{BS}} \rangle^{}_{\cal S}$ we obtain
\begin{eqnarray}
\label{eqmintegral}
\lefteqn{ ^{\Gamma I}\!\!{\cal AM}^{b}_{\mu}(q,E_d + {\textstyle \frac{3}{4}}
 q^2) =
\, ^{\Gamma I}\!\!  {\cal AB}_{\mu}^b(q,E_d + {\textstyle \frac{3}{4}} q^2)}
\\
&+ & \sum_{b'} \sum_{\nu \rho} \int\limits_{0}^{\infty} \! dq' \, q'^2 \;
{^{\Gamma I}\!\!{\cal A V}}_{\mu \nu}^{b b'}(q, q', E_d +
{\textstyle \frac{3}{4}} q^2) \, \nonumber \\
& & \hspace{8mm} \times \, \Delta^{\eta '}_{\nu \rho}
(E_d + {\textstyle \frac{3}{4}} q^2 - {\textstyle \frac{3}{4}} q'^2)
\, ^{\Gamma I} \!\!{\cal AM}_{\rho}^{b'}(q',E_d +
{\textstyle \frac{3}{4}} q^2) \,. \nonumber
\end{eqnarray}
\noindent
with
\begin{eqnarray}
 ^{\Gamma I} \!\!{\cal AM}_\mu^b (q,E) && \\
& & \hspace{-1.6cm} = {\frac{1}{\sqrt 3}} \sum_{ \beta}
\langle{(\beta) \, g_\mu \, q \, b \, \Gamma; I} |
{G_0 (E + i0) \Omega_\beta^{(-) \dagger} \,
 H_{\rm em}}|{\Psi_{\text {BS}}} \rangle^{}_{\cal S} \,, \nonumber 
\end{eqnarray}
\noindent
and
\begin{eqnarray}
  ^{\Gamma I}\!\!{\cal AB}_\mu^b (q, E) && \\
& & \hspace{-1.6cm} =  {\frac{1}{\sqrt 3}} \sum_{ \beta}
  \langle{(\beta)\, g_\mu \, q\, g \, b \,  \Gamma; I }
  |{G_0 (E + i\,0) H_{\rm em}}|{ \Psi_{\text{BS}}} \rangle^{}_{\cal S} \,,
\nonumber
\end{eqnarray}
\noindent
where we have used the separable expansion of Eq. (\ref{eqtmatest})
for the $T$ operator.  Here $^{\Gamma I} \!\!{\cal AB}_\mu^b$
represents the so-called plane-wave (Born) amplitude and $^{\Gamma
I} \!\!{\cal AM}_\mu^b$ denotes the full final-state amplitude. The
effective potential $^{\Gamma I} \!\!{\cal AV}_{\mu \nu}^{bb'}$
entering Eq. (\ref{eqmintegral}) is given by
\begin{eqnarray}
 ^{\Gamma I}\!\!{\cal AV}^{b b'}_{\mu \nu} (q,E) &=& \sum_\beta
(1 -  \delta_{\beta \gamma}) \, \\
& & \hspace{-1.2cm} \times \,\langle {(\beta) \, g_\mu \, q \, b \, \Gamma; I} |
{G_0 (E + i0)}|
{ \, g_\nu \, q' \, b' \, \Gamma'; I' (\gamma)}\rangle \,. \nonumber
\end{eqnarray}
\noindent
Noting that the two non--zero contributions to the sum are identical,
we can write the effective potential as

\begin{eqnarray}
\label{effpot}
 ^{\Gamma I}\!\!{\cal A V}^{b b'}_{\mu \nu} (q, q', E) & =&
2 \sum_{l l'}
\int\limits_{0}^{\infty} \int\limits_{0}^{\infty} dp \, p^2 \, dp' \,
p'^2 \, g^{\eta}_{l \mu}(p) \, \\
& & \hspace{-2.3cm} \times \, \langle (\beta) \, p \, q\, l\, b\, \Gamma; I
 | G_0(E + i0) | p' \,q'\, l'\, b'\, \Gamma; I \,(\alpha)\rangle \,
g^{\eta'}_{l' \nu}(p'). \nonumber
\end{eqnarray}
\noindent
The recoupling coefficients entering this equation can be found in
Ref. \cite{Januschke93} (or in a more compact form in
\cite{Gloeckle83} for another coupling sequence which can easily be
changed to the present one). In
Eqs. (\ref{eqmintegral})-(\ref{effpot}) we have used the fact that the
Born term, the effective potential, and therefore the full amplitude
are diagonal in the quantum numbers $\Gamma$, $M_\Gamma$, $I$, and
$M_I$.

In order to be able to solve Eq. (\ref{eqmintegral}) numerically an
off-shell extension is required. This can easily be achieved by
replacing $E_d + \frac{3}{4} q^2$ with the energy parameter $E$.  The
solution of Eq. (\ref{eqmintegral}) is obtained on the real axis by
expanding the solution in cubic $B$ splines \cite{deBoor78} and
solving a system of linear equations for the unknown coefficients.
The logarithmic singularities in the kernel of the integral equation
have been treated by a standard subtraction technique. For more
details on the numerical solution of an integral equation of similar
type we refer to \cite{Januschke93} and references therein.

In the Born amplitude one finds that the terms in the summation are
independent from the partition, i.e., the summation over the different
clusters can be replaced by a factor of 3. Using the states of
Eq. (\ref{eqpartwavegeneral}) we obtain

\begin{eqnarray}
 ^{\Gamma I}\!\!{\cal AB}^b_\mu (q, E) &=&
  {\sqrt 3}  \sum_l  \int \! dp \, p^2  \,
  \frac{  g^{\eta}_{l \mu} (p)}
  {E - p^2 -\frac{3}{4} q^2 + i \,0} \, \nonumber \\
& & \hspace{1.5cm} \times \, \langle{ p \, q \,  l\, b\, \Gamma; I}
|{H_{\rm em}}|
 {\Psi_{\text{BS}}} \rangle^{}_{\cal S} \,.\
\end{eqnarray}

\noindent
For energies $E$ above the deuteron breakup we have to take care of
the pole in the propagator, which is done by using a standard
subtraction technique.

The three-nucleon bound state $| \Psi_{\text{BS}} \rangle $ with
binding energy $E_{BS}$ which is contained in the expression for the
Born amplitude is determined by the Schr{\"o}dinger equation
\vspace{-2mm}
\begin{equation}
\label{eqeigen}
(E_{\text{BS}} - H)  \, |\Psi_{\text{BS}}\rangle = 0,
\end{equation}
\noindent
where the total Hamiltonian $H$ is given by $H = H_0 + V = H_0 +
\sum\limits_{\gamma = 1}^3 V_\gamma$.  When introducing the channel
resolvents $G_\gamma(z) = (z - H_0 - V_\gamma)^{-1}$,
Eq. (\ref{eqeigen}) can be written in form of a homogeneous integral
equation,
\begin{eqnarray}
|{\Psi_{\text{BS}}}\rangle &=& G_{\gamma}(E_{\text{BS}}) \,
\overline{V}_{\gamma} \,
|{\Psi_{\text{BS}}} \rangle  \nonumber \\
& = & G_{\gamma}(E_{\text{BS}}) \sum_\beta (1 -
\delta_{ \gamma \beta})\, {V_{\beta}}\, |{\Psi_{\text{BS}}} \rangle,
\end{eqnarray}
\noindent
with $\overline{V}_{\gamma} = V - V_{\gamma}$ being the channel
interaction between particle $\gamma$ and the $(\alpha \beta$)
subsystem.  If we now introduce the position $ |{F_\beta}\rangle = (V
- V_\beta) \; |{\Psi_{\text{BS}}}\rangle$, and use the relation
$V_\gamma G_\gamma = T_\gamma G_0$, we obtain the equation
\begin{equation}
\label{eqform}
 |{F_\beta}\rangle = \sum_\gamma (1 - \delta_{\beta \gamma})\,
 T_\gamma(E_{\text {BS}}) \,G_0(E_{\text {BS}})\,| {F_\gamma} \rangle ,
\end{equation}
\noindent
where $G_0$ again is the resolvent of the free Hamiltonian.  In
Eq. (\ref{eqform}), the summation runs over all two-fragment
partitions $\gamma$.  The ``form-factors'' $|{F_\beta}\rangle$ are
related to $|{\Psi_{\text{BS}}} \rangle$ by
\begin{eqnarray}
 |{\Psi_{\text {BS}}}\rangle &=& \sum_\gamma G_0(E_{\text {BS}}) \,
 T_\gamma(E_{\text {BS}}) \, G_0(E_{\text {BS}}) \, |{F_\gamma} \rangle
\nonumber \\
&=&
 \sum_\gamma |{\psi_\gamma}\rangle \,,
\label{psi}
\end{eqnarray}
\begin{minipage}[t]{8.3cm}
\begin{table}[ht]
\begin{center}
\begin{tabular}{ccc}
 Bonn {\em A} (EST) & Bonn {\em B} (EST)  & Paris (EST) \\
\hline
 -8.284 & -8.088 & -7.3688
\end{tabular}

\vspace{3mm}
\caption{\label{energytab} Calculated binding energies for the three\-nucleon
 bound state.
The total angular momentum of the two-body subsystem was
restricted to $j\leq 2$.}
\end{center}
\vspace{-0.8cm}
\end{table}
\end{minipage}

\noindent
where the $|{\psi_\gamma}\rangle $ are the standard Faddeev
components.  For a numerical treatment of Eq. (\ref{eqform}) we
multiply this equation with $G_0$ and the partial-wave states
$\langle p \, q \,l\, b\, \Gamma; I |$.  After inserting the separable
$T$ matrix from Eq. (\ref{eqtmatest}) and defining

\begin{equation}
 F^{\mu b}_\beta (q) = \sum_l \int\limits_{0}^{\infty} dp \, p^2 \,
g^{\eta}_{l \mu} (p) \,
 \langle p \, q \, l \, b\, \Gamma; I | G_0 | F_\beta \rangle,
\end{equation}

\noindent
Eq.~(\ref{eqform}) goes over into

\begin{eqnarray}
\label{form}
  F^{\mu b} (q) &=& \sum_{b'} \sum_{\nu \rho} \int\limits_{0}^{\infty} dq'
 \, q'^2 \, {\cal A V}^{b b'}_{\mu \nu} (q, q', E_{\text{BS}}) \, \\
& & \hspace{1.5cm} \times \, \Delta^{\eta'}_{\nu \rho}
(E_{\text{BS}} - {\textstyle \frac{3}{4}} q'^2)  \, F^{\rho b'} (q') .
\nonumber
\end{eqnarray}

\noindent
After discretization Eq. (\ref{form}) can be treated as a linear
eigenvalue problem, where the energy is considered as a parameter
which is varied until the corresponding eigenvalue equals unity. The
eigenvalues can be found by using standard numerical algorithms.  The
obtained binding energies for the three potentials used in this paper
can be found in Tab. \ref{energytab}. As shown in \cite{Schadow2000a}
these values are practically the same as those for the original
potentials.

The whole wave function can now be calculated by either using Eq.
(\ref{psi}), or by applying the permutation operator $P$ on one
Faddeev component \cite{Gloeckle83}

\begin{equation}
\label{eqperm}
| \Psi_{\text{BS}} \rangle = (1 + P) \, | \psi_1 \rangle,
\end{equation}

\noindent
where $P$ represents the sum of all cyclical and anticyclical
permutations of the nucleons. The required antisymmetry of the wave
function is achieved by projecting only on those channels with
$(-)^{l+s+t} = -1$.

In our calculation of the Faddeev components the total angular
momentum $j$ of the two-body potential was restricted to $j \leq 2$
(18 channels), while in the full state all partial waves with $j \leq
4$ (34 channels) have been taken into account. With this number of
channels converged calculations of the observables for the
photoprocesses in this paper were achieved, incorporating 99.8\% of
the wave functions \cite{Schadow98a,Sandhas98a}. For more details
concerning the properties of the wave functions and their high quality
we refer to \cite{Schadow2000a}.

The relevant electromagnetic operator in the total cross section at
low energies is a dipole operator. In the differential cross section
and at higher energies also the quadrupole operator is relevant.  In
order to take into account meson exchange currents we use Siegert's
theorem \cite{Siegert37}, then these operators are given by
\cite{Gianini85}

\begin{equation}
{ {H_{\rm em}^{(1)}}} = - {\cal N}  \sqrt{\frac{4 \pi}{3}} i \, E_\gamma\,
\sum_{i = 1}^3 e_i  \, r_i \, Y_{1 \lambda}(\vartheta_i, \varphi_i)
\end{equation}

\begin{figure}[hbt]
\centerline{\psfig{file=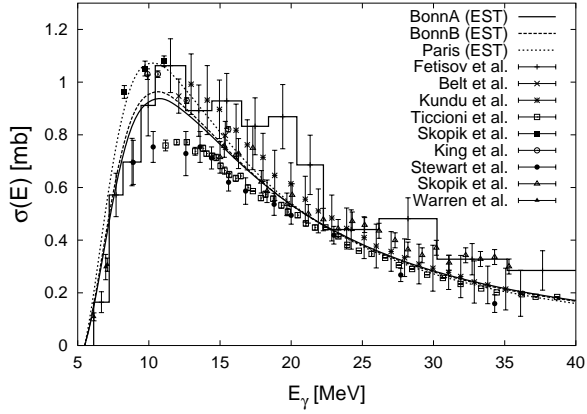,width=8cm,angle=-90}}
\vspace{2mm}
\caption{\label{fig3hetot}Total cross section for the
photodisintegration of $^3$He.  The data are from [2-4,8-10,12-14,20].}
\end{figure}

\noindent
and

\begin{equation}
{ {H_{\rm em}^{(2)}}} = {\cal N} \sqrt{\frac{4 \pi}{3}} \frac{E_\gamma^2}
{\sqrt{20}}\,
\sum_{i = 1}^3 e_i  \, r_i^2 \, Y_{2 \lambda}(\vartheta_i, \varphi_i) ,
\end{equation}

\noindent
where $E_{\gamma}$ denotes the photon energy. Here, $r_i$ are the
nucleon coordinates, $e_i$ the electric charges, and $\lambda = \pm 1$
is the polarization of the photon. The normalization factor ${\cal N}$
contains the quantization volume and is canceled out in the
calculation of the cross sections. For the calculation of matrix
elements of these operators with initial and final state they need to
be transformed into the three-body center--of--mass
system. Expressions for the matrix elements can be found in
\cite{Schadow97c}.

The on-shell amplitudes can be obtained from

\begin{eqnarray}
\label{eqampl}
\lefteqn{^{(-)}_{\;\;\;\cal S}\langle \, {\bf q} \,S \, M_S; \psi_d j m_j \,
| \,  H_{\rm em} \,|\,  \Psi_{\text {BS}} \, \Gamma'
M_{\Gamma'} \,\rangle^{}_{\cal S}} \\
 & =  &
 \sum_{\Gamma = \frac{1}{2}}^{\frac{5}{2}}
    \sum_{M_\Gamma} \sum_{b}  \sum_{M_K M_L} \,
Y_{L M_L} ( \hat q) \,
  \langle \, j m_j S M_S \,|\, K M_K \, \rangle
\nonumber  \\
& & \times \; \langle \, K M_K L M_L \,|\, \Gamma M_{\Gamma} \, \rangle
\,  {^{\Gamma I}}\! {\cal A M}_1^{b}\;
   (q, E_d + \textstyle \frac{3}{4}\, q^2) \,. \nonumber
\end{eqnarray}

\vspace{2mm}
\begin{figure}[hbt]
\centerline{\psfig{file=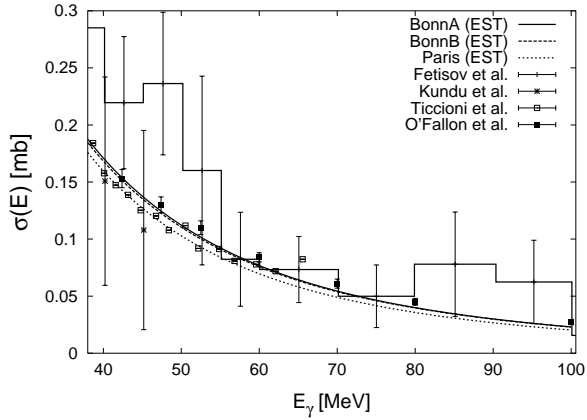,width=8cm,angle=-90}}
\vspace{2mm}
\caption{\label{fig3hetot-100}Total cross section for the
photodisintegration of $^3$He.  The data are from [4,9,10,16].}
\end{figure}

\begin{figure}[hbt]
\centerline{\psfig{file=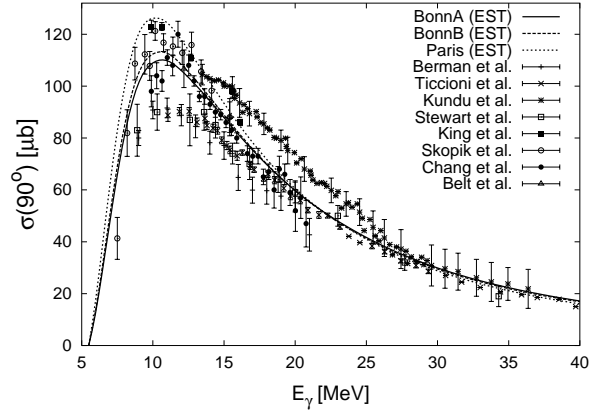,width=8cm,angle=-90}}
\vspace{2mm}
\caption{\label{fig3he-diff90}Differential cross section at 90$^{\rm o}$ for the
 photodisintegration of $^3$He. The  data are from [2,5,8-13].}
\end{figure}

\noindent
In the summation of course only those channels contribute that have a
deuteron as their subsystem.  The total angular momentum of the
bound-state wave functions is $\Gamma = \frac{1}{2}$. This means when
taking into account $E2$ contributions, the maximum total angular
momentum of the outgoing state can be $\Gamma = \frac{5}{2}$.  With
these amplitudes the unpolarized differential cross section for the
photodisintegration process is given by

\begin{eqnarray}
\label{wq2}
\lefteqn{ \frac{d\sigma}{d\Omega}(q, \theta) = m_N \hbar \,
\frac{q E_\gamma}{3 \pi c} (2 \pi)^3
  \frac{1}{4}} \\
& & \;\; \times \!\!\!\!\sum_{M_S M_J} \sum_{\lambda M_\Gamma'}
  \left |  ^{(-)}_{\;\;\;\cal S}\langle \, {\bf q} \, S \,M_S; \psi_d j m_j \,
 | \, H_{\rm em} \,|\,\Psi_{\text {BS}} \,\Gamma'
M_{\Gamma'} \,\rangle^{}_{\cal S}\right |^2 \,. \nonumber
\end{eqnarray}

\noindent
The differential cross section is usually expanded in terms of Legendre
polynomials

\begin{equation}
\label{eqlegendre}
\sigma(q,\theta) = \frac{d\sigma}{d\Omega}(q, \theta) = A_0 \left ( 1 + \sum_{k = 1}^{4} a_k \, P_k (\cos \theta)
\right ).
\end{equation}

\noindent
The coefficients $A_0$ and $a_k$ can be calculated analytically from
Eqs. (\ref{eqampl}) and (\ref{wq2}).  The total cross section is
obtained by integrating Eq. (\ref{eqlegendre}) over the angle $\theta$
between

\begin{figure}[hbt]
\centerline{\psfig{file=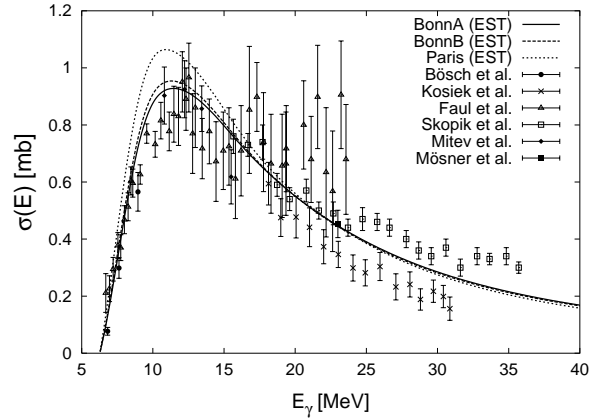,width=8cm,angle=-90}}
\vspace{2mm}
\caption{
\label{fig3htot}
Total cross section for the photodisintegration of $^3$H.
The  data are from [22-28].}
\end{figure}

\begin{minipage}[ht]{17.5cm}
\begin{table*}[ht]
\begin{center}
\begin{tabular}{ccccccc}
$E_x$ [MeV] & $A_0$ [$\mu b$]& $a_1$ & $a_2$ & $a_3$ & $a_4$ \\
\hline
21.47  & 0.575\phantom{$\pm$ 0.3} & 0.42\phantom{$\pm$0.1} &
-0.83\phantom{$\pm$0.13} & -0.41\phantom{$\pm$0.15} &
-0.05\phantom{$\pm$0.1} \\
 & 0.51$\pm$ 0.03 & 0.28$\pm$0.1 & -0.82$\pm$0.13 & -0.38$\pm$0.15 &
-0.07$\pm$0.1 \\
24.14  & 0.517\phantom{$\pm$ 0.3} & 0.44\phantom{$\pm$0.1} &
-0.79\phantom{$\pm$0.13} & -0.43\phantom{$\pm$0.15} &
-0.05\phantom{$\pm$0.1} \\
& 0.42$\pm$ 0.04 & 0.34$\pm$0.1 & -0.86$\pm$0.12 & -0.35$\pm$0.15 &
-0.12$\pm$0.1 \\
26.81   & 0.467\phantom{$\pm$ 0.3} & 0.46\phantom{$\pm$0.1} &
-0.75\phantom{$\pm$0.13} & -0.44\phantom{$\pm$0.15} &
-0.06\phantom{$\pm$0.1} \\
& 0.38$\pm$ 0.03 & 0.41$\pm$0.1 & -0.86$\pm$0.13 & -0.39$\pm$0.15 &
-0.11$\pm$0.1 \\
29.47  & 0.425\phantom{$\pm$ 0.3} & 0.47\phantom{$\pm$0.1} &
-0.72\phantom{$\pm$0.13} & -0.45\phantom{$\pm$0.15} &
-0.06\phantom{$\pm$0.1} \\
& 0.33$\pm$ 0.03 & 0.39$\pm$0.1 & -0.77$\pm$0.10 & -0.42$\pm$0.15 &
-0.12$\pm$0.1 \\
32.14  & 0.389\phantom{$\pm$ 0.3} & 0.48\phantom{$\pm$0.1} &
-0.68\phantom{$\pm$0.13} & -0.45\phantom{$\pm$0.15} &
-0.07\phantom{$\pm$0.1} \\
& 0.33$\pm$ 0.03 & 0.33$\pm$0.1 & -0.79$\pm$0.12 & -0.36$\pm$0.15 &
-0.08$\pm$0.1
\end{tabular}

\vspace{3mm}
\caption{\label{tablegendre} Coefficients for the expansion of the
differential cross section for $p$-$d$ capture. For each energy the
first row corresponds to the theoretical results obtained with the
Paris (EST) potential, and the second row contains the data from
Ref. [19].}
\end{center}
\vspace{-1cm}
\end{table*}
\end{minipage}

\noindent
 the incoming photon and the outgoing proton or neutron

\begin{equation}
\label{eqtotal}
\sigma  = 4 \pi \, A_0.
\end{equation}

The cross section for the $p$-$d$ or $n$-$d$ capture process is
obtained from the corresponding photodisintegration expression by
using the principle of detailed balance \cite{Craver77a}

\begin{equation}
\frac{d \sigma^{\rm dis}}{d \Omega} =
\frac{3}{2} \, \frac{k^2}{Q^2} \,
\frac{d \sigma^{\rm cap}}{d \Omega}.
\end{equation}

\noindent
Here, $k$ and $Q$ are the momenta of the proton and the photon,
respectively. In the present treatment no Coulomb forces have been
taken into account, in other words the matrix elements of
Eq. (\ref{bampl}) for $p$-$d$ capture differ from the corresponding
$n$-$d$ expression only in their isospin content.

\section{Results}
\label{secresults}

It should be pointed out that we have shifted all theoretical cross
sections to the experimental threshold for a meaningful comparison.
All calculations are done with the theoretical binding energies.

\begin{figure}[hbt]
\centerline{\psfig{file=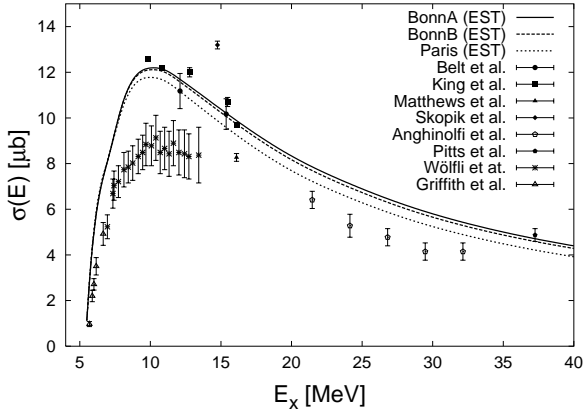,width=8cm,angle=-90}}
\vspace{2mm}
\caption{
\label{figcap3htot}
Total cross section for the capture of protons by deuterons.
The  data are from [1,6-8,12,13,18,19,21].}
\end{figure}

$ $
\vspace{6.3cm}

In Fig. \ref{fig3hetot} we show our theoretical results for the total
cross section of $^3$He photodisintegration compared to most of the
available experimental data
\cite{Stewart63,Warren63a,Fetisov65a,Belt70,Kundu71a,Ticcioni73,Chang74,Skopik79a,King84a}
up to $E_\gamma$~=~40~MeV.  Not shown are the data by van der Woude
{\em et al.} \cite{Vanderwoude71a} and Chang {\em et
al}. \cite{Chang72a} who found evidence for an excited state in their
measurements. This resonance behavior has never been confirmed by any
other group. It can be seen from Fig. \ref{fig3hetot} that there are
large discrepancies between the different data sets around
$E_\gamma$~=~11~MeV. The theoretical curves lie in between the data
sets.  It should be emphasized that for the calculated curves there is
a correlation between the three-body binding energy and the peak
height of the cross section for the photodisintegration
\cite{Schadow98a,Sandhas98a}, i.e., the higher the binding energy the
lower the cross section at the peak. Above 12 MeV the mentioned
discrepancy of the experimental data declines. Due to the large error
bars it is not possible to draw further conclusions.

In Fig. \ref{fig3hetot-100} we present total cross section
calculations of photodisintegration of $^3$He between $E_{\gamma} =
40$ MeV and $E_{\gamma} = 100$ MeV compared to the measurements of
Fetisov {\em et al}. \cite{Fetisov65a}, Kundu {\em et al}.
\cite{Kundu71a}, Ticcioni {\em et al}. \cite{Ticcioni73}, and O'Fallon
{\em et al}.  \cite{OFallon72a}. For energies above $E_{\gamma} = 60$
MeV the measured points lie slightly above our curves, computed by
employing Bonn {\em A}, Bonn {\em B}, and Paris potentials. However,
our curves agree with the tendency of the data.

We would like to point out that especially in this energy range a high
rank representation of the $NN$ poten-

\vspace{4mm}

\begin{minipage}[h]{8.3cm}
{\begin{table}[ht]
\begin{center}
\begin{tabular}{ccccccc}
$E_x$ [MeV] & $A_0$ [$\mu b$]& $a_1$ & $a_2$ & $a_3$ & $a_4$ \\
\hline
21.47 & 0.579& -0.071&   -0.88 &   0.069 &  -0.0021 \\
24.14 & 0.517& -0.065&   -0.84 &   0.064 &  -0.0023 \\
26.81 & 0.464& -0.057&   -0.81 &   0.056 &  -0.0026 \\
29.47 & 0.421& -0.047&   -0.78 &   0.047 &  -0.0028 \\
32.14 & 0.383& -0.035&   -0.75 &   0.037 &  -0.0030 \\
\end{tabular}
\vspace{3mm}
\caption{\label{tablegendre2} Coefficients for the expansion of the
differential cross section for $n$-$d$ capture for the
Paris (EST) potential.}
\vspace{-1.4cm}
\end{center}
\end{table}
}
\end{minipage}

\twocolumn[
\hsize\textwidth\columnwidth\hsize\csname @twocolumnfalse\endcsname

{\begin{figure}[hbt]
\centerline{\hbox{
\psfig{figure=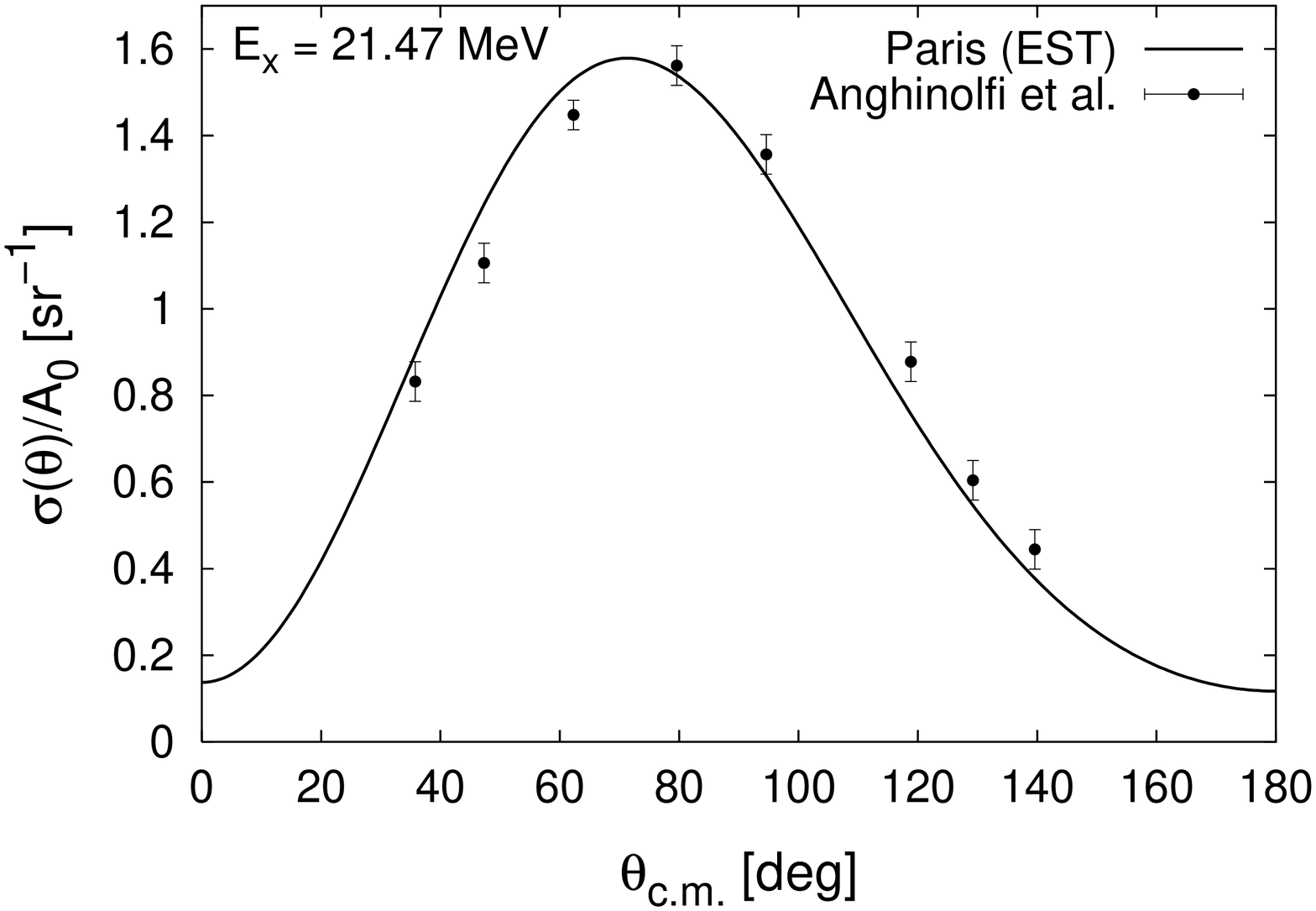,width=80mm,angle=-90}
\hspace{0.5cm}
\psfig{figure=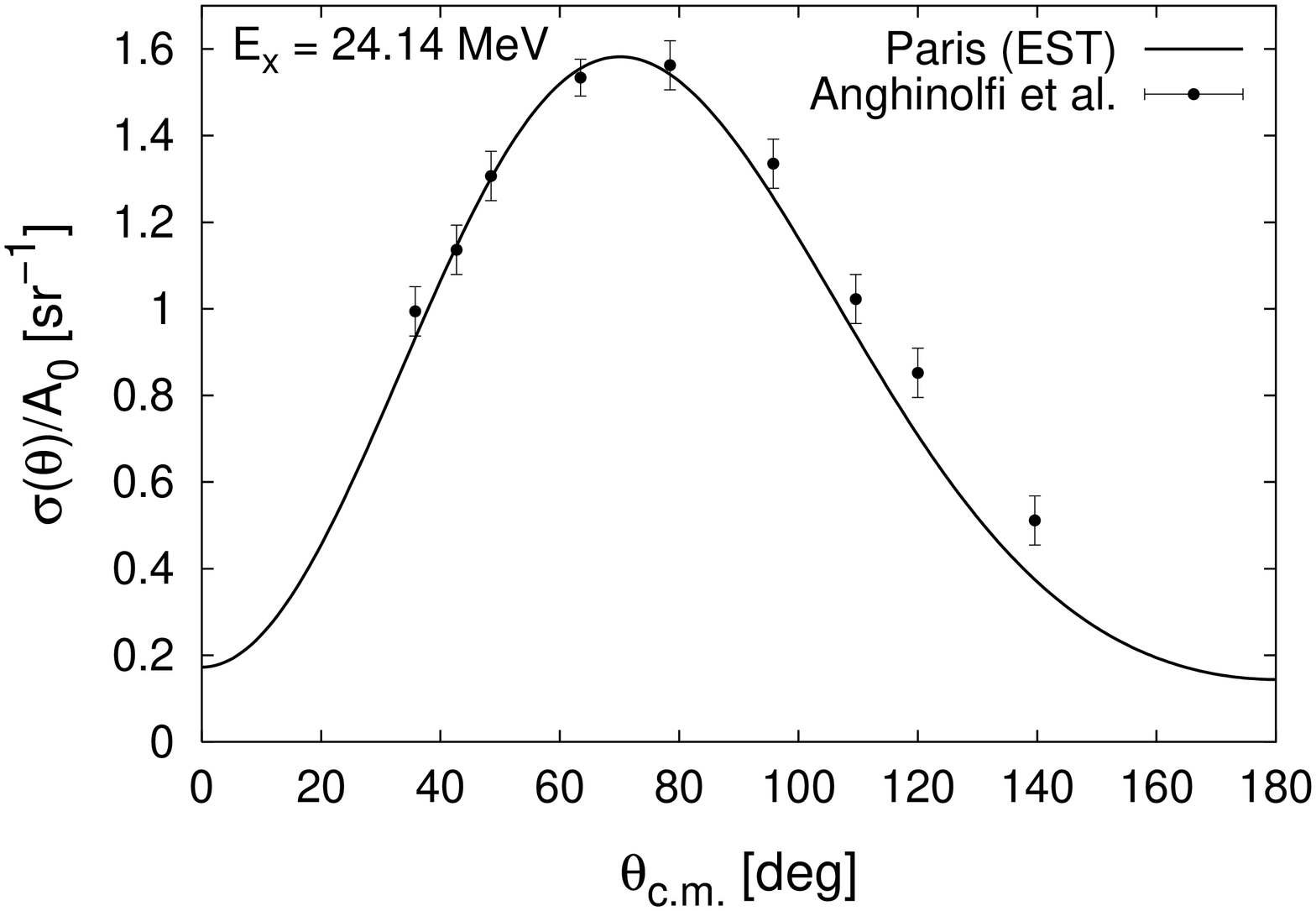,width=80mm,angle=-90}
}}
\centerline{\hbox{
\psfig{figure=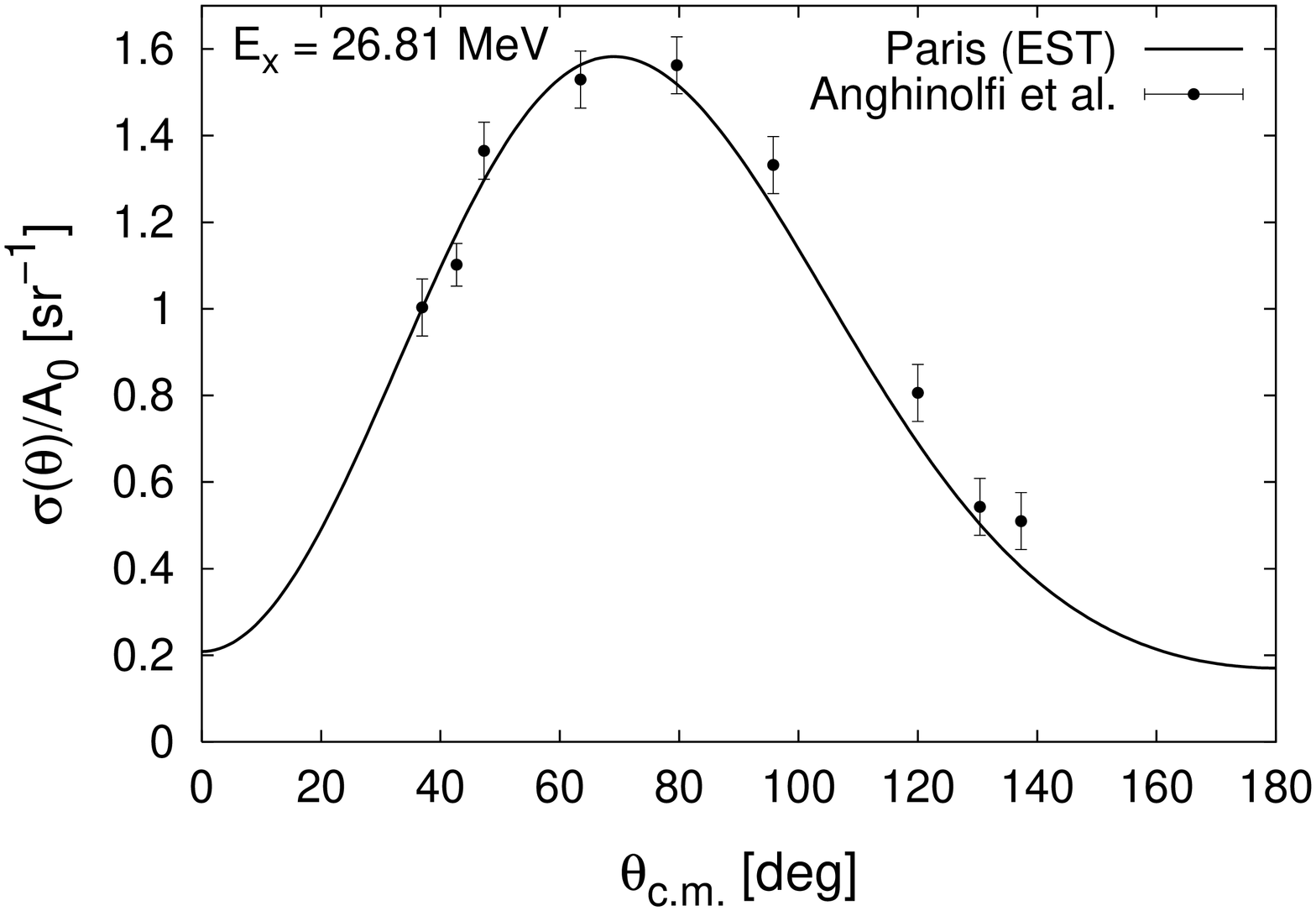,width=80mm,angle=-90}
\hspace{0.5cm}
\psfig{figure=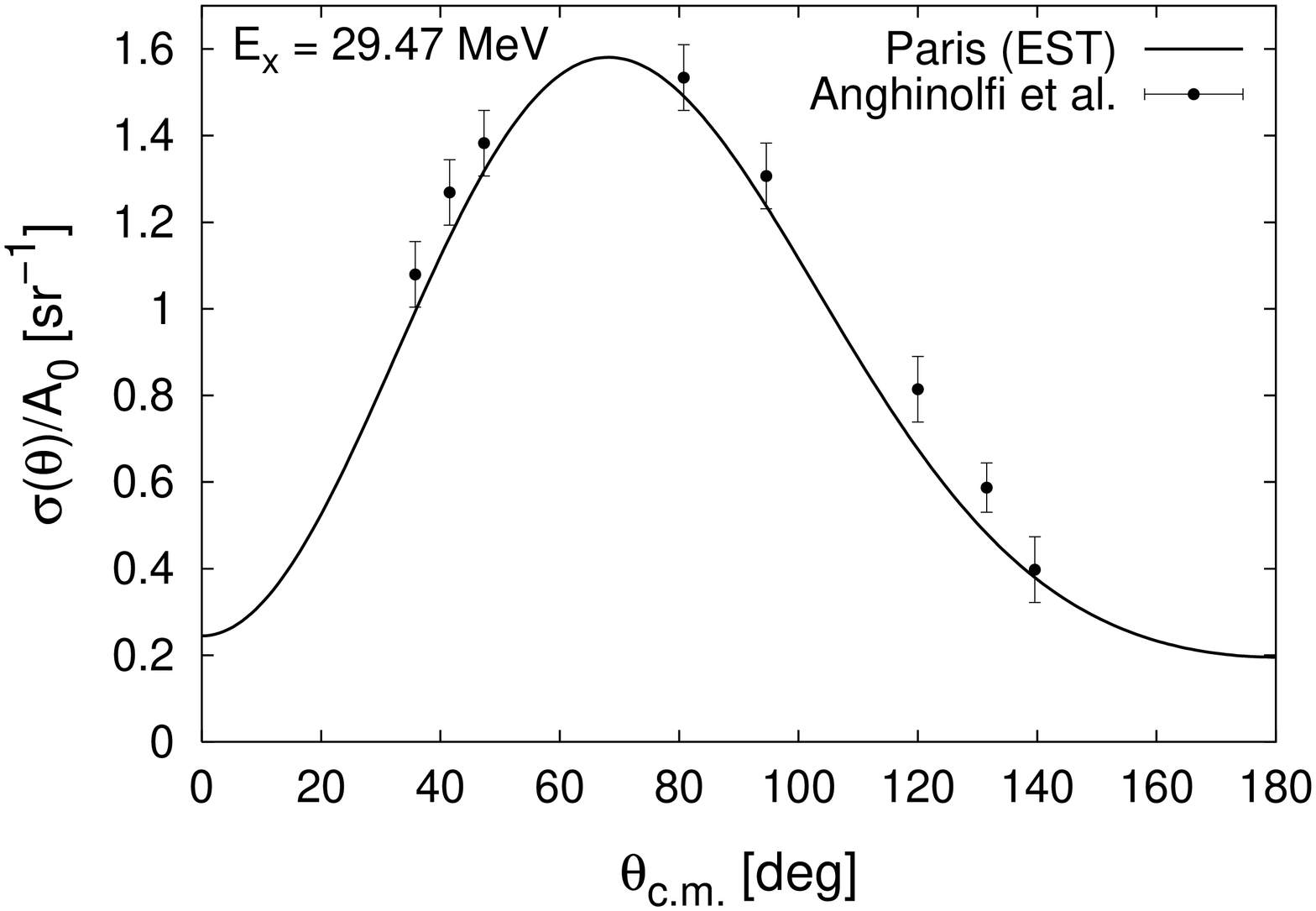,width=80mm,angle=-90}
}}
\centerline{\hbox{
\psfig{figure=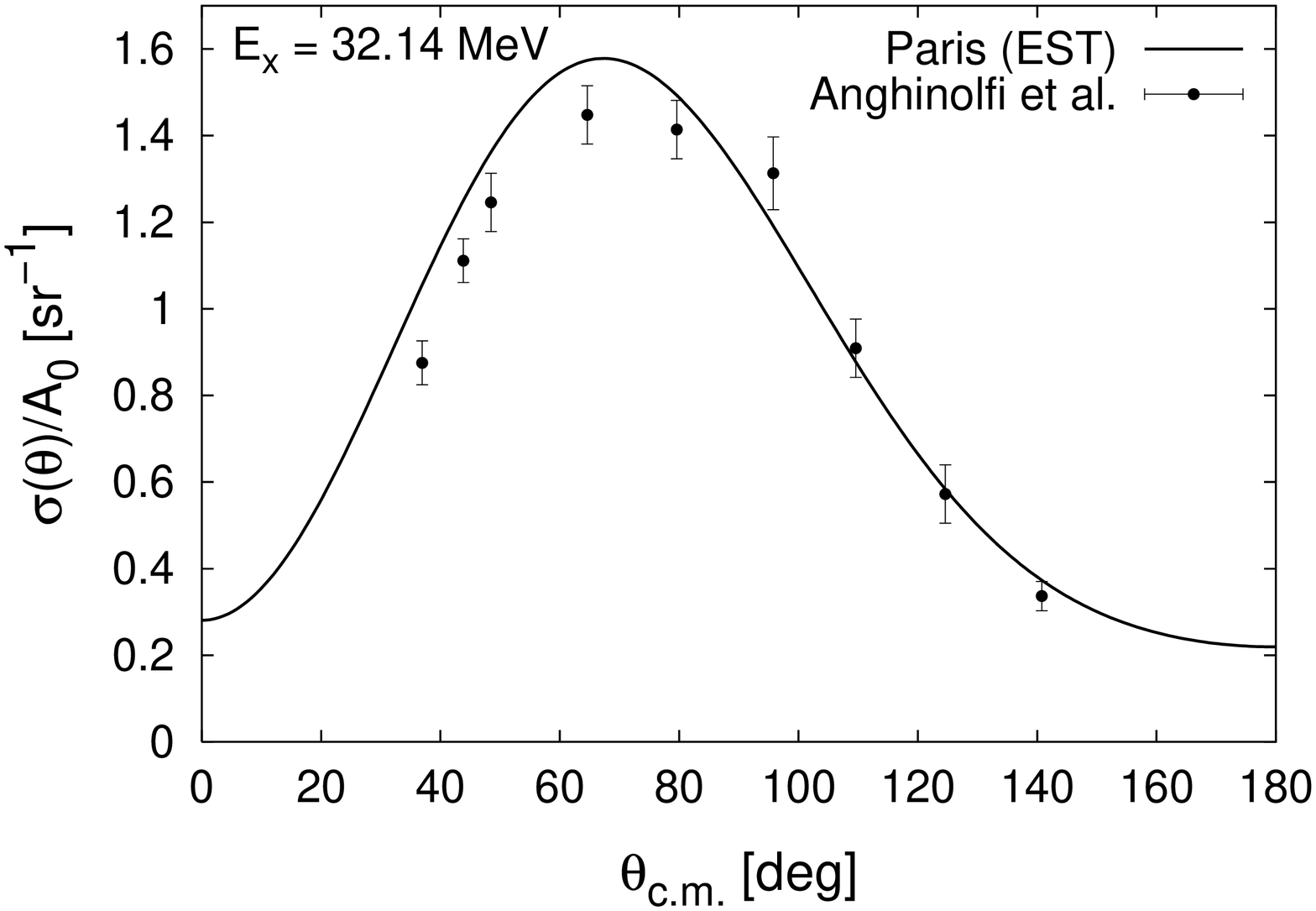,width=80mm,angle=-90}
\hspace{0.5cm}
\psfig{figure=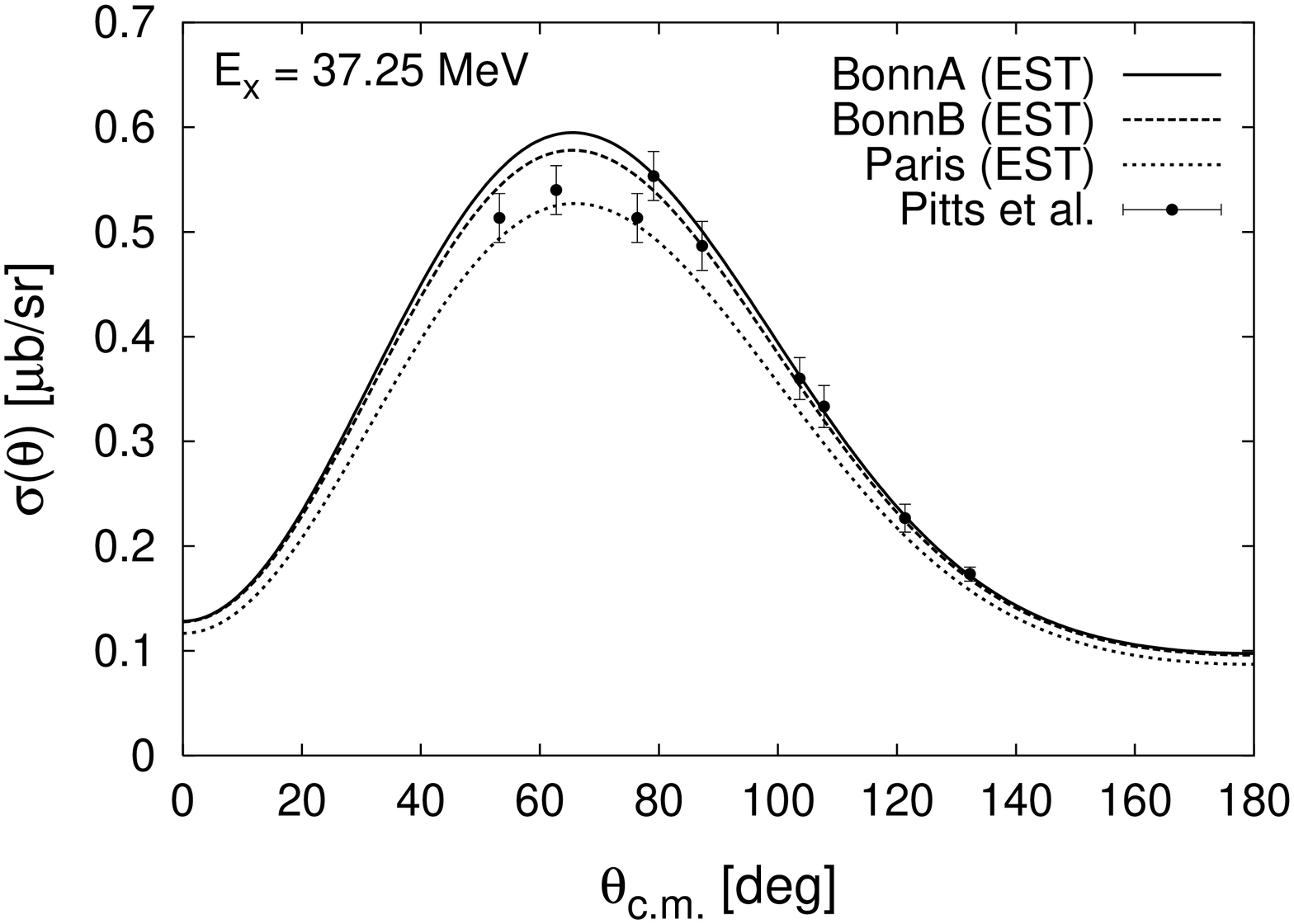,width=80mm,angle=-90}
}}
\vspace{2mm}
\caption{
\label{fig3he-diff}
Angular distribution and differential cross section for the capture of
protons by deuterons. The data are from [19,21].}
\vspace{-2mm}
\end{figure}
}
]
\noindent
tials is required in order to get converged results. Below $E_\gamma$
= 40 MeV the improvements with respect to a low rank calculation are of the
order of 1-5\%.  In view of the experimental errorbars this change is
of course not relevant. Above $E_\gamma$ = 40 MeV the low rank
calculations yield a cross section which is 5-15\% lower than the high
rank calculations presented in this paper.

The differential cross section calculations at 90$^{\rm o}$ for the
photodisintegration of $^3$He up to an energy of $E_{\gamma} = 40$ MeV
are illustrated in Fig. \ref{fig3he-diff90} in comparison to the
corresponding experimental data
\cite{Stewart63,Berman64a,Belt70,Kundu71a,Ticcioni73,Chang74,Skopik79a,King84a}.
The EST representations of the potentials Bonn {\em A}, Bonn {\em B},
and Paris are employed. As in the previous figures, we can not observe
any significant difference for energies above $E_{\gamma} = 20$ MeV,
whereas the peak region shows a considerable potential dependence, as
discussed for Fig. \ref{fig3hetot}. The data by Berman {\em et
al}. \cite{Berman64a} are below the calculated curves, however, they
do agree with the tendency in the peak region. There is a remarkable
discrepancy between these data sets and the data points measured by
Kundu {\em et al}. \cite{Kundu71a}, but they coincide with the
theoretical curves for energies above 25 MeV. We find again that the
theoretical results lie in between the data, though there are
discrepancies between the data sets \hfill and, \hfill moreover, the
\hfill errorbars \hfill are

\twocolumn[
\hsize\textwidth\columnwidth\hsize\csname @twocolumnfalse\endcsname
{
\begin{figure}[hbt]
\centerline{\hbox{
\psfig{figure=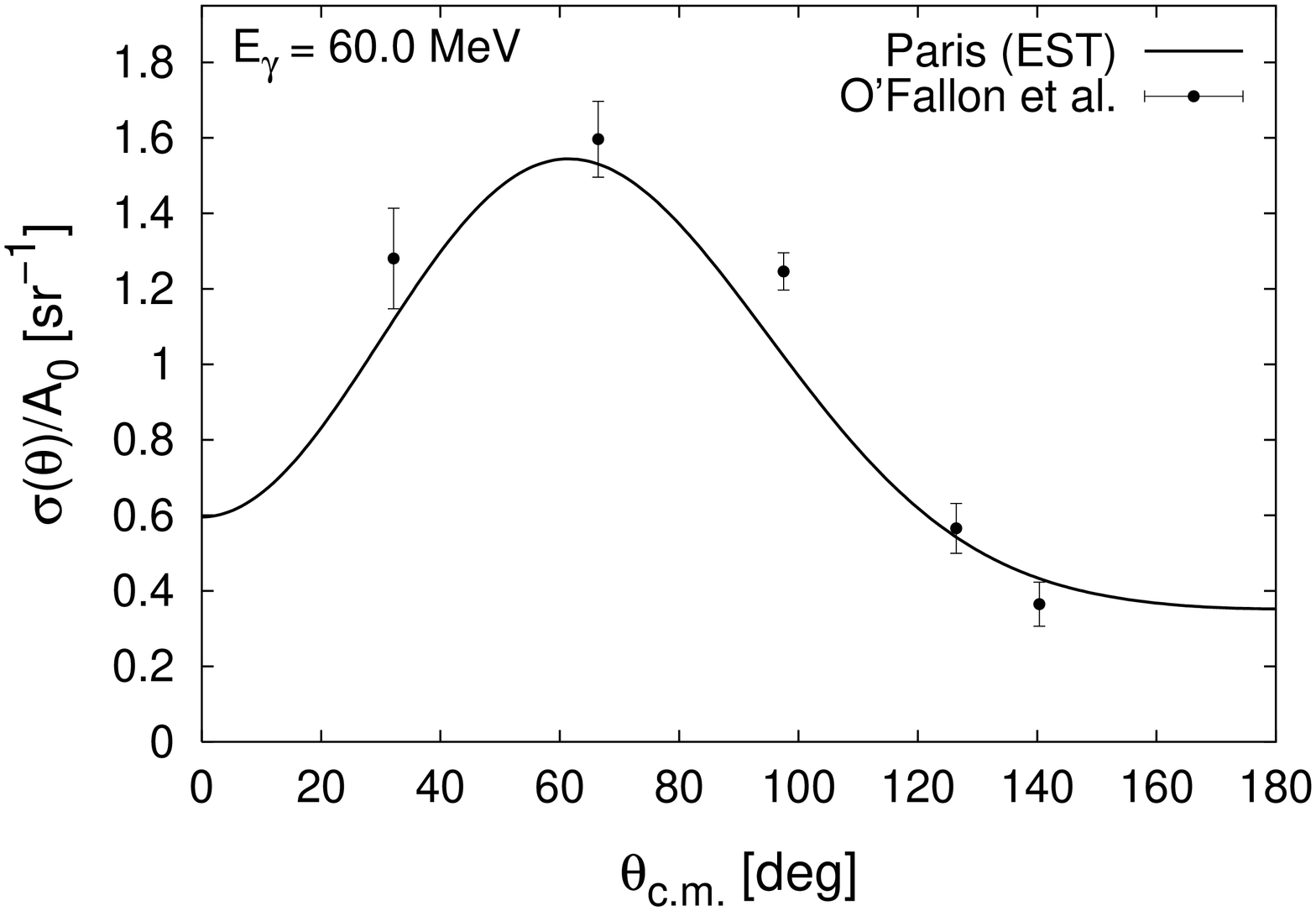,width=80mm,angle=-90}
\hspace{0.5cm}
\psfig{figure=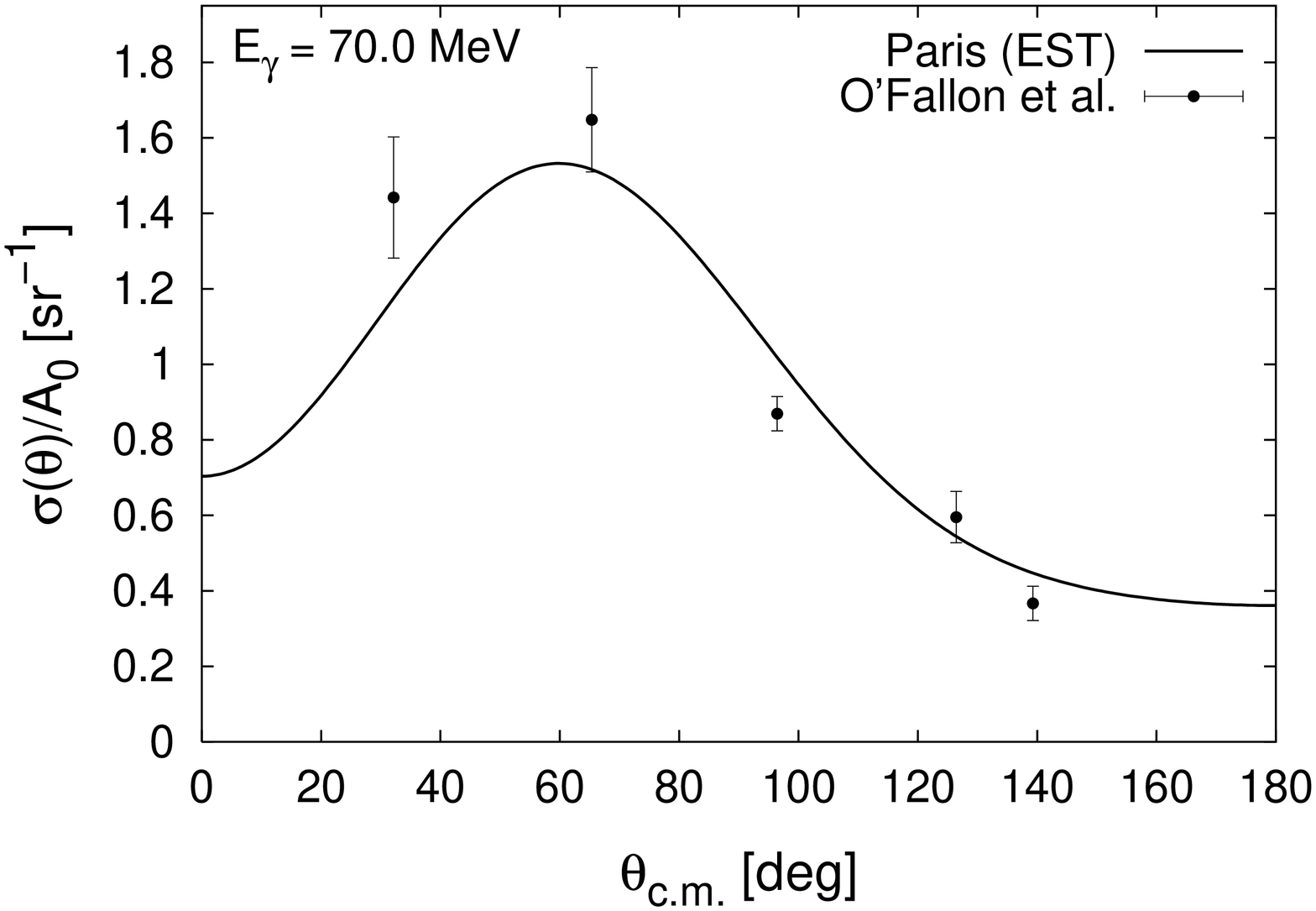,width=80mm,angle=-90}
}}
\centerline{\hbox{
\psfig{figure=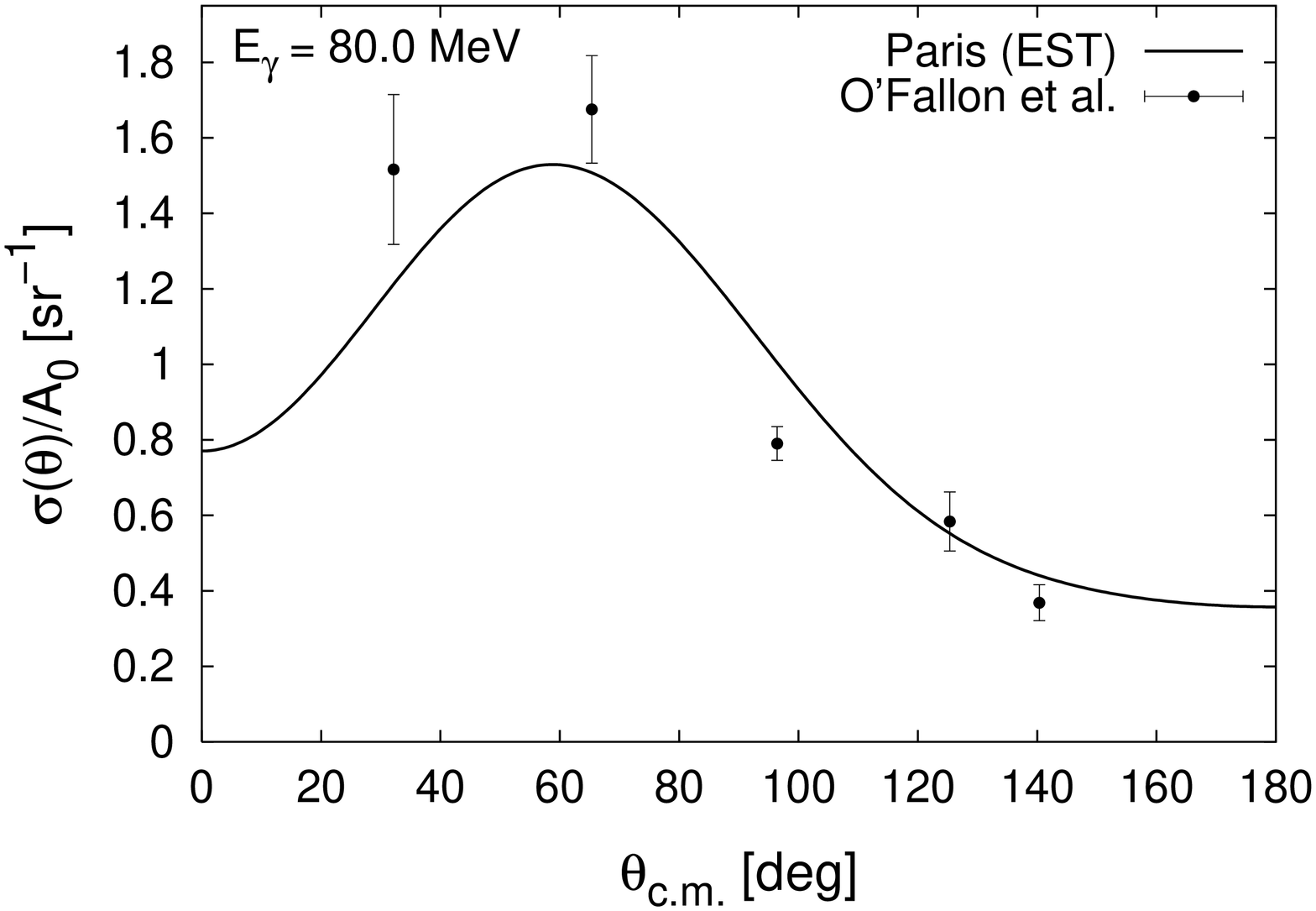,width=80mm,angle=-90}
\hspace{0.5cm}
\psfig{figure=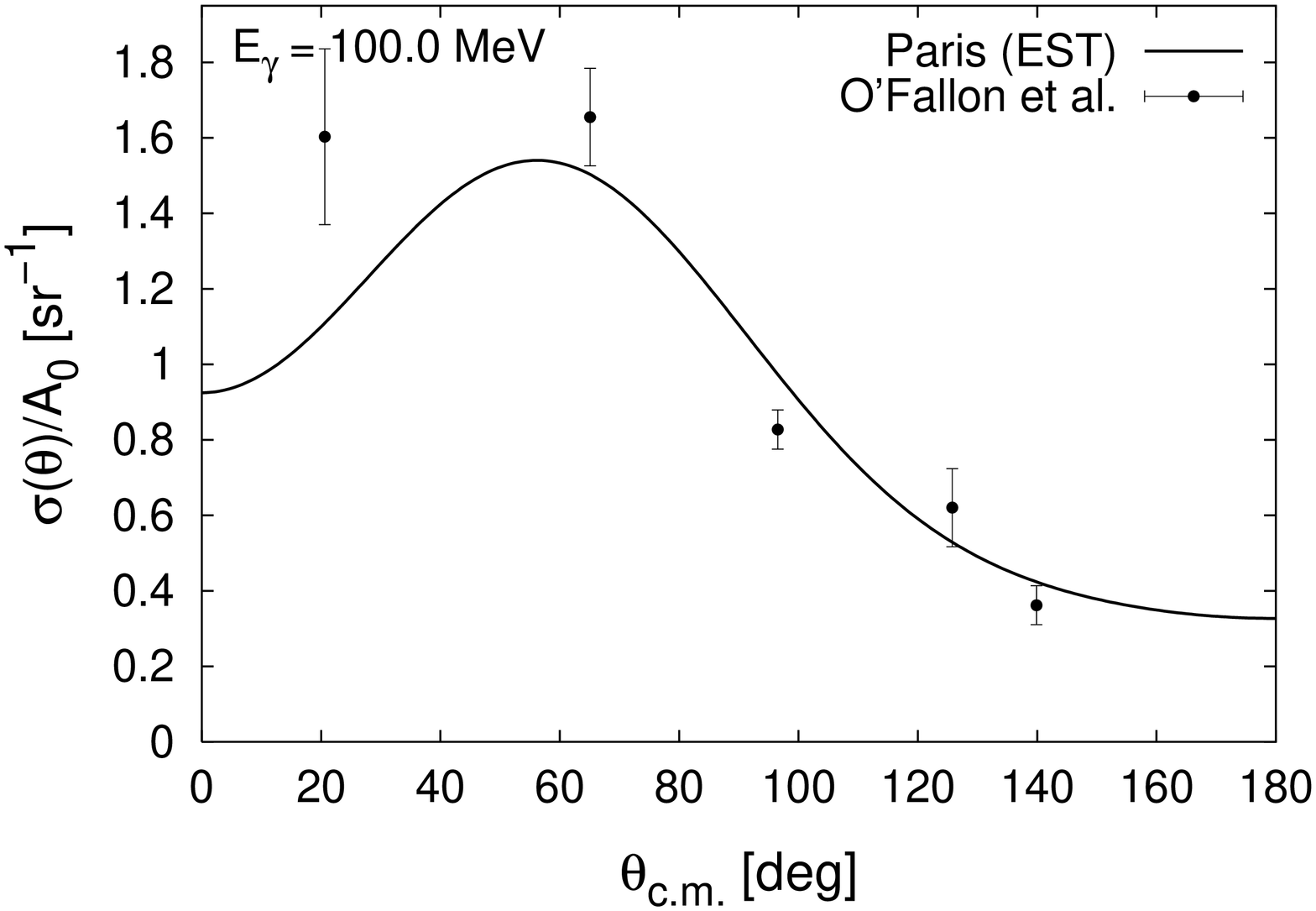,width=80mm,angle=-90}
}}
\vspace{2mm}
\caption{
\label{fig3he-diff2}
Angular distribution and differential cross section for the
photodisintegration $^3$He. The data are from [16].}
\end{figure}

}
]

\noindent
quite large.

The two-body photodisintegration of $^3$H has been measured by
B{\"o}sch {\em et al}. \cite{Boesch64}, Kosiek {\em et al}.
\cite{Kosiek66,Pfeiffer68}, Faul {\em et al}.  \cite{Faul80}, and
Skopik {\em et al}. \cite{Skopik81}.  In Fig. \ref{fig3htot} we
display these data compared to our theoretical calculations.  Also
shown in this figure are the transformed results by Mitev {\em et
al}. \cite{Mitev86a} and M{\"o}sner {\em et al}. \cite{Moesner86a}
obtained from radiative $n$-$d$ capture measurements.  It should be
pointed out that the most recent measurement by M{\"o}sner {\em et
al}. is in excellent agreement with the theory.  We notice that for
low energies, e.g., up to $E_\gamma=15$ MeV, the calculated curves for
the different potentials show a different behavior, whereas for higher
energies all three calculations do not yield any significant
difference. In the peak region, the curves for the Bonn {\em A} and
the Bonn {\em B} potentials cover the experimental data in between the
error bars better. For energies above 20 MeV there is a large
discrepancy between the data sets by Kosiek {\em et al}.  and Skopik
{\em et al}., although the tendency of the data is similar. This
indicates a normalization problem.

In Ref. \cite{Schadow99a} we discussed the available data for $p$-$d$
capture below $E_x = 20$ MeV. It was shown, that only the coefficient
$A_0$ of the expansion in Eq. (\ref{eqlegendre}) has some potential
dependence, whereas the coefficients $a_k$ are almost independent from
the interaction. Also in this case there is a correlation between the
peak height and the binding energy, i.e., the lower the binding energy
the lower the peak height. It should be pointed out, that this is the
inverse of the relation found in case of the photodisintegration. We
also have demonstrated that there seems to be a normalization problem
in the experimental data.  It can be seen in Fig. \ref{figcap3htot}
that the data by W{\"o}lfli {\em et al}. \cite{Woelfli66,Woelfli67},
Matthews {\em et al}.  \cite{Matthews74} and Anghinolfi {\em et
al}. \cite{Anghinolfi83a} are too low compared to those by King {\em
et al}. \cite{King84a}, Pitts {\em et al}. \cite{Pitts88a}, and Belt
{\em et al}. \cite{Belt70} which agree with our theoretical
curves. This indicates a calibration problem of the measurements.  It
was also shown in \cite{Schadow99a} that after renormalization the
data sets by Matthews {\em et al}. are in agreement with those by King
{\em et al}. and the theoretical curves. At energies above $E_\gamma$
= 20 MeV we encounter a similar problem and compare in
Fig. \ref{fig3he-diff} the differential cross section divided by
$A_0$.  It can be seen that the agreement between theory and the
experimental data by Anghinolfi {\em et al}. is very good.  A
comparison of the expansion coefficients obtained by Anghinolfi {\em
et al}.  and our theoretical values for the Paris (EST) potential is
given in Tab. \ref{tablegendre}. There are discrepancies for $A_0$
which are connected to the normalization problem mentioned
earlier. Despite the relatively big experimental error bars for the
expansion coefficients $a_k$ there are considerable good agreements.

Also shown in Fig. \ref{fig3he-diff} are the data by Pitts {\em et
al}. \cite{Pitts88a}.

\begin{figure}[hbt]
\centerline{\psfig{file=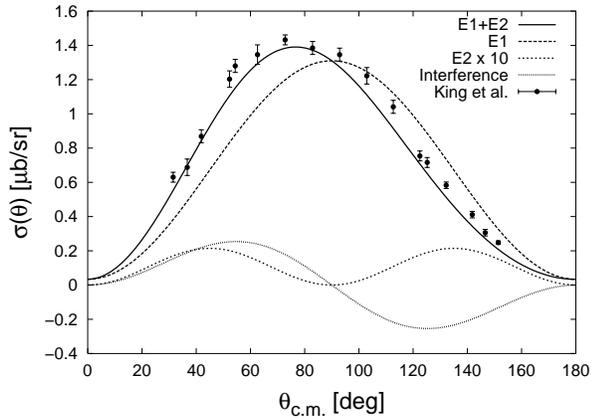,width=8cm,angle=-90}}
\vspace{2mm}
\caption{
\label{fig3he-e1e2}
Differential cross section for the capture of protons by
deuterons at $E^{\rm lab}_p$ = 10.93 MeV for the Bonn {\em B} potential. The
data are from [13].}
\end{figure}

\noindent
In this case there is also excellent agreement for the absolute cross
section, particularly by employing the Bonn {\em A} potential. There
are two additional data sets by van der Woude {\em et al.}
\cite{Vanderwoude71a} at $E_x$ = 19.2 MeV and $E_x$ = 20.6 MeV which
are not shown here because of the measured resonance behavior, as
mentioned above.

Besides total cross section data, O'Fallon {\em et al}.
\cite{OFallon72a} have also measured data sets of the differential cross
section for photodisintegration of $^3$He up to an energy of
$E_\gamma$ = 140 MeV.  As can be seen in Fig. \ref{fig3hetot-100}
their total cross sections are slightly higher than the theoretical
predictions. For a meaningful comparison with our calculations we illustrate
in Fig. \ref{fig3he-diff2} four of their data sets for the differential
cross section normalized with $A_0$. Within their error
bars they agree quite well with the theoretical calculations.

The different contributions for the $E1$ and $E2$ transitions in case
of $p$-$d$ capture at $E_p^{\rm lab} = 10.93$ MeV are shown in
Fig. \ref{fig3he-e1e2}. The pure $E2$ contributions are very small and
enter the differential cross section essentially through the $E1$-$E2$
interference term, which leads to the
asymmetry, i.e., the curve is \hfill shifted to \hfill smaller \hfill angles.

\vspace{1mm}

\begin{figure}[hbt]
\centerline{\psfig{file=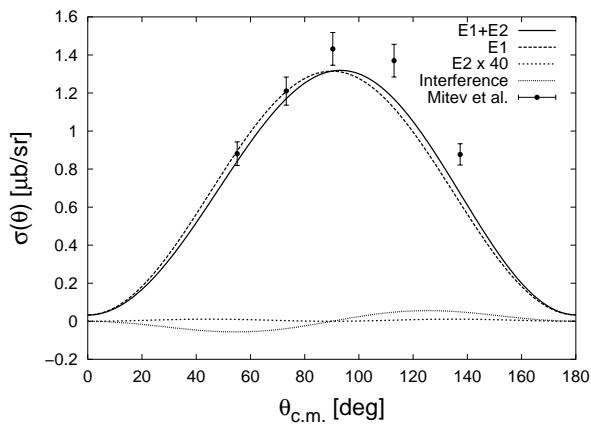,width=8cm,angle=-90}}
\vspace{2mm}
\caption{
\label{fig3h-e1e2}
Differential cross section for the capture of neutrons by
deuterons at $E^{\rm lab}_n$ = 10.8 MeV for the Bonn {\em B} potential. The
data are from [27].}
\end{figure}

\begin{figure}[hbt]
\centerline{\psfig{file=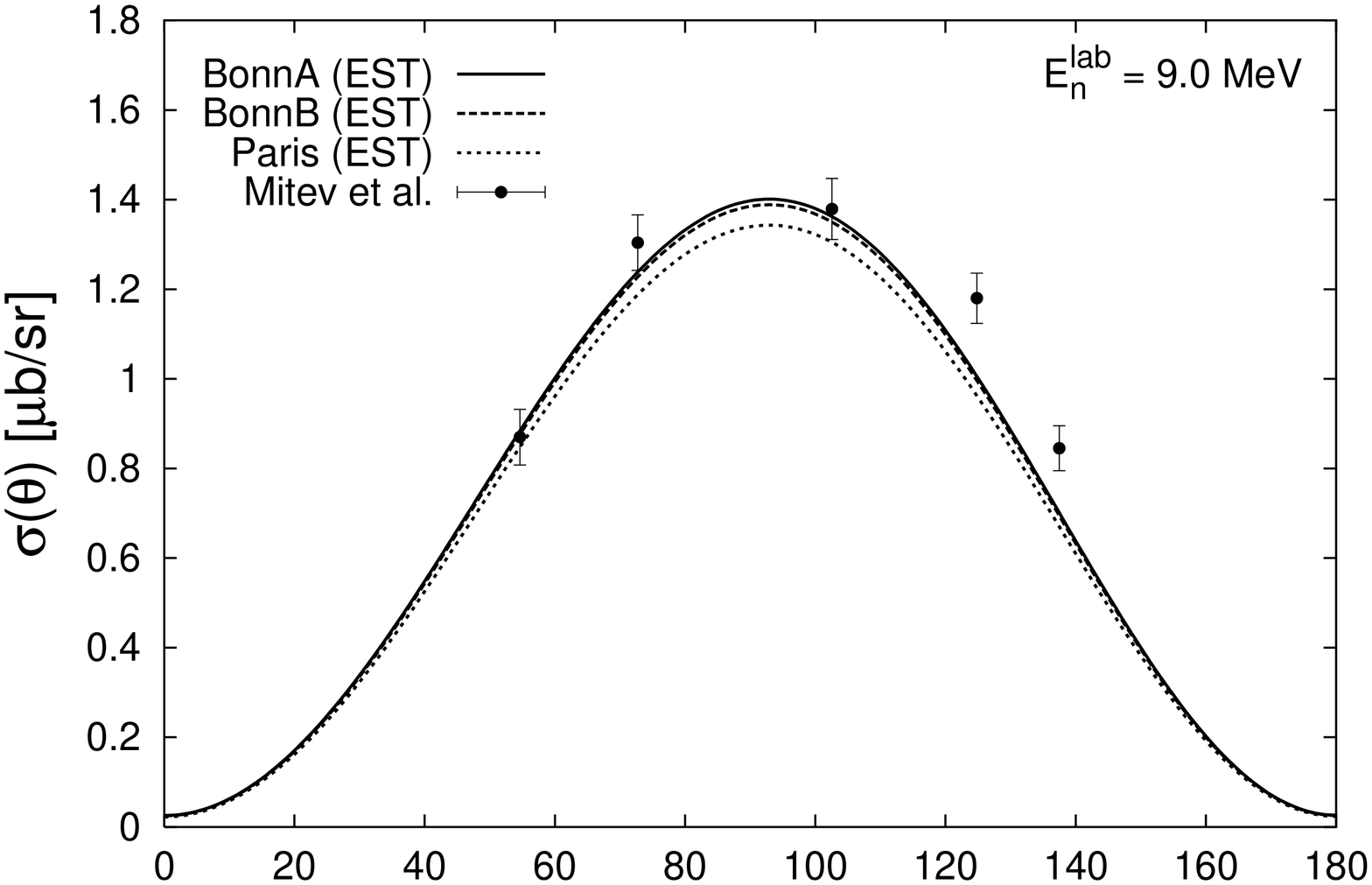,width=8cm,angle=-90}}
\vspace{-6mm}
\centerline{\psfig{file=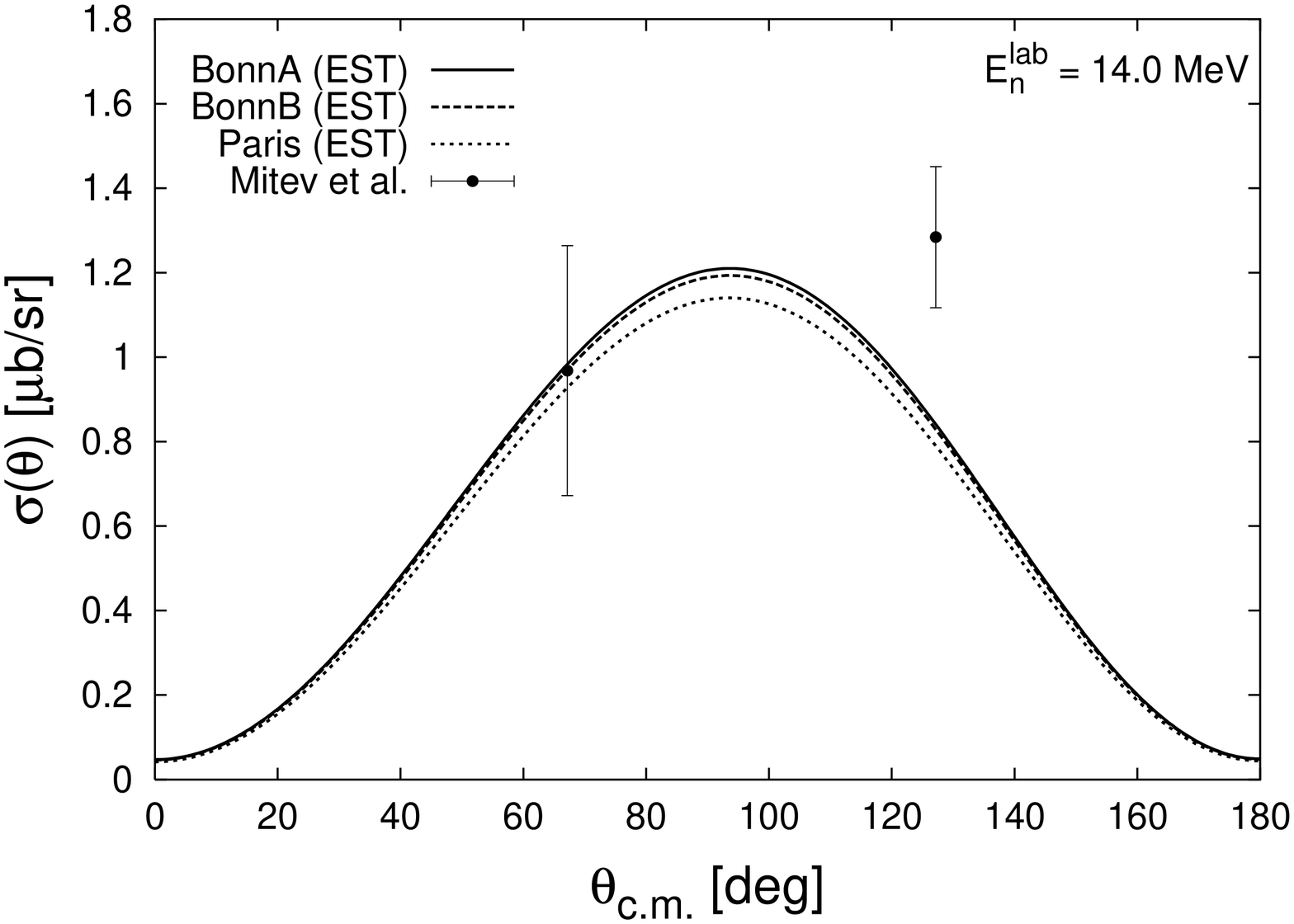,width=8cm,angle=-90}}
\vspace{2mm}
\caption{
\label{fig3hdiffcross}
Differential cross section for the capture of neutrons by
deuterons. The  data are from [27].}
\end{figure}

\noindent
 With the inclusion of $E2$ transitions there is an excellent
agreement with the data by King {\em et al}. \cite{King84a}.  The only
measurement of the angular distribution of the differential cross
section for $n$-$d$ capture has been done by Mitev {\em et
al}. \cite{Mitev86a}.  The different contributions for the $E1$ and
$E2$ terms is shown in Fig. \ref{fig3h-e1e2} for $E_n^{\rm lab} =
10.8$ MeV. Due to isospin selection rules the $E2$ contribution, and
hence the interference between the $E1$ and the $E2$ term, is much
smaller compared to $^3$He. It should be noted that in this case the
maximum of the differential cross section is shifted to larger
angles. This observation was also made in Ref. \cite{Klepacki88b} for
the Born approximation.  A comparison to the theoretical calculations
at $E_n^{\rm lab} = 9$~MeV and $E_n^{\rm lab} = 14$ MeV is shown in
Fig. \ref{fig3hdiffcross}. It can be seen that the peaks of the
experimental data tend to have a bigger asymmetry than the theoretical
curves. It is remarkable that this circumstance is independent of the
potential choice. This indicates either a stronger contribution of a
higher multipole, not present in our theoretical calculations, or an
error in the data. To the best of our knowledge there are no other
differential cross section data for $^3$H photodisintegration or the
inverse reaction available.  There are also no experimental data
available for the Legendre coefficients $a_k$.  Nevertheless, we show
for comparison in Tab. \ref{tablegendre2} corresponding calculated
values for $n$-$d$ capture at the same energies as the available data
for $p$-$d$ capture \hfill from Tab. \ref{tablegendre}. It \hfill can
be \hfill seen \hfill that for $n$-$d$ capture \hfill the \hfill
angular \hfill distribution is dominated by

\begin{figure}[hbt]
\centerline{\psfig{file=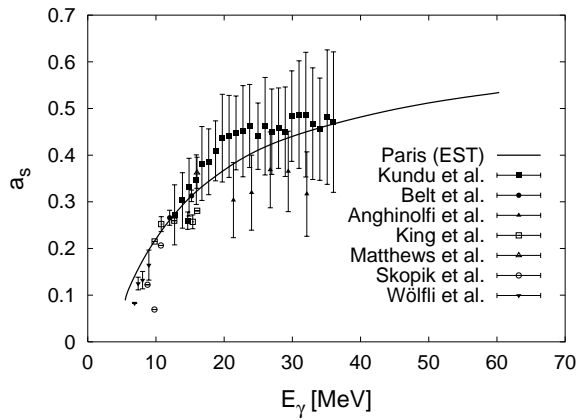,width=8cm,angle=-90}}
\vspace{2mm}
\caption{
\label{fig3he-as}
Fore-aft asymmetry $a_s$ for $^3$He as defined in the text. The
data are from [6-9,12,13,18-20].}
\end{figure}

\noindent
 the $a_2$
coefficient, i.e., the $E1$ transitions, whereas in case of $p$-$d$
capture there are bigger admixtures of $E2$ contributions.

One observable of particular interest in the context of angular
distributions is the so-called fore-aft asymmetry. This quantity is
defined by

\begin{equation}
a_s = \frac{\sigma(54.7^{\rm{o}}) - \sigma(125.3^{\rm{o}})}
{\sigma(54.7^{\rm{o}}) + \sigma(125.3^{\rm{o}})} \,.
\end{equation}

\noindent
In terms of Legendre polynomial expansion coefficients of Eq.
(\ref{eqlegendre}), this can be written as

\begin{equation}
a_s = \frac{a_1 - \frac{2}{3} a_3}{\sqrt{3} (1 - \frac{7}{18} a_4)} \,.
\end{equation}

\begin{figure}[hbt]
\centerline{\psfig{file=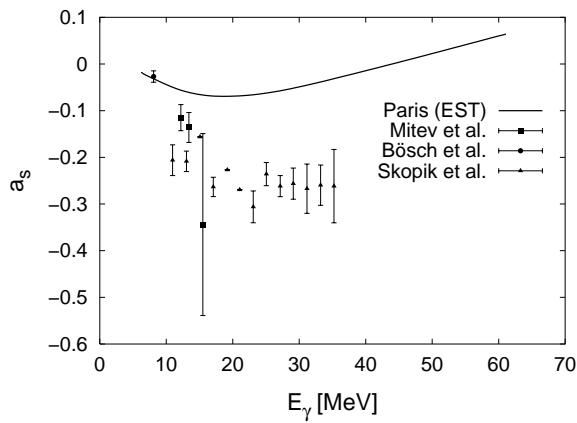,width=8cm,angle=-90}}
\vspace{2mm}
\caption{
\label{fig3h-as}
Same as Fig. \ref{fig3he-as} but for $^3$H. The data are from [22,26,27].}
\end{figure}

\noindent
In Figs. \ref{fig3he-as} and \ref{fig3h-as} we compare our results
with the available data sets. The theoretical curves for $^3$He agree
quite well with the data
\cite{Woelfli66,Woelfli67,Belt70,Kundu71a,Skopik79a,King84a,Matthews74,Anghinolfi83a,Skopik83a},
whereas the calculated asymmetry for $^3$H is smaller by a factor 5
than the experimental data \cite{Boesch64,Skopik81,Mitev86a}. A
similar observation was made by Skopik {\em et al}.  \cite{Skopik83a}
using their effective capture calculations, where no FSI effects were
taken into account. Since all experimental data show a consistently
higher fore-aft asymmetry, there seems to be \hfill something

\noindent
missing in the theoretical description of this reaction. One possible
explanation for the discrepancy between theory and experiment was that
the FSI has not been taken into account properly in previous
calculations. In the present calculations we have shown that this
discrepancy still remains when taking FSI effects into account. Therefore, it
is still unclear where the differences stem from. The M1 term is not
likely to solve the problem, since it is only expected to have an
effect at extreme angles or at very low energies. Also three-nucleon
forces are not expected to solve the problem since the angular
distribution shows no potential dependence. A possible solution could
be the inclusion of explicit meson exchange currents which allow a
stronger coupling of higher multipoles to $^3$H.

\vspace{-3mm}

\section{Conclusions}
\label{secconclusions}

\vspace{-1mm}

In this paper we have analyzed all available experimental data for the
photodisintegration of $^3$He and $^3$H and the corresponding inverse
reactions below $E_\gamma = 100$ MeV by comparing with our
calculations using realistic $NN$ interactions.  We have shown that
the theoretical curves agree with the experimental data for the total
cross section within the error bars. Moreover, in many cases the
measured differential cross sections for $p$-$d$ capture (aside from a
normalization factor) can be explained theoretically over a large
energy range. In \cite{Schadow99a} it was already shown that a similar
normalization problem exists for the data below $E_x$ = 20 MeV. There,
it was also shown that the angular distribution is insensitive to the
underlying two-body interaction, whereas there is a strong correlation
between the three-body binding energy and the normalization constant
$A_0$ \cite{Schadow98a,Sandhas98a}.  Since the angular distribution is
insensitive to the employed interaction, we do not expect large
effects of three-nucleon forces. On the other hand taking account of
them will change the three-body binding energy and hence the
normalization constant $A_0$.

For $n$-$d$ capture the description of the angular distribution is
less good. For energies above 10 MeV the theoretical results give a
much smaller asymmetry than the experimental data. Hence, the
theoretical fore-aft asymmetry shows a large discrepancy from the
experimental data, whereas for $p$-$d$ we achieve a very good
agreement.

\vspace{-1mm}

\acknowledgements
\vspace{-1mm}
The work of W. Sch. is supported by the Natural Science and
Engineering Research Council of Canada. The work of O. N.  is
supported by the Graduiertenkolleg "Die Erforschung subnuklearer
Strukturen der Materie".  W. Sch. acknowledges support and hospitality
from the University of Padova during the write up of this paper. We
thank L. Canton for providing computing time, which made the high rank
calculations of this work possible.

\end{document}